\documentclass{emulateapj}
\usepackage{graphicx}
\usepackage{amssymb}
\usepackage{amsmath}
\usepackage{ulem}
\usepackage[colorlinks=true,linkcolor=blue,anchorcolor=blue,citecolor=blue,urlcolor=blue]{hyperref}  

\shorttitle{The Structure of the Strongly Lensed Gamma-ray Source B2~0218+35}
\shortauthors{Barnacka et al.}

\begin{document}

\title{The Structure of the Strongly Lensed Gamma-ray Source B2~0218+35}

\author{Anna Barnacka$^{1,2}$}
\author{Margaret J. Geller$^1$}
\author{Ian P. Dell'Antonio$^3$}
\author{Adi Zitrin$^4$}
\affil{$^1$Harvard-Smithsonian Center for Astrophysics, 60 Garden St, MS-20, Cambridge, MA 02138, USA\\
$^2$Astronomical Observatory, Jagiellonian University, Cracow, Poland \\
$^3$Department of Physics, Brown University, Box 1843, Providence, RI 02912\\
$^4$Cahill Center for Astronomy and Astrophysics, California Institute of Technology, MC 249-17, Pasadena, CA 91125, USA}

\email{abarnacka@cfa.harvard.edu}

\begin{abstract}

Strong gravitational lensing is a powerful tool for resolving the high energy universe. 
We combine the temporal resolution of {\it Fermi}-LAT,
the angular resolution of radio telescopes,
and the independently and precisely known Hubble constant from the analysis by the Planck collaboration,
to resolve the spatial origin of  gamma-ray flares in the strongly  lensed source B2~0218+35.
The lensing model achieves $1\,$milliarcsecond  spatial resolution of the source at gamma-ray energies.
The data imply that the gamma-ray flaring sites are separate from  the radio core:
the bright gamma-ray flare (MJD: 56160 - 56280) occurred $51\pm8\,$pc from the 15~GHz radio core, toward the central engine. 
This displacement is significant at the $\sim3\sigma$ level, and is limited primarily by the precision of the Hubble constant. 
B2~0218+35 is the first source where the position of the gamma-ray emitting region relative to the radio core can be resolved. 
We discuss the potential of  an ensemble of strongly lensed high energy sources for elucidating the physics of distant variable sources based on data from {\it Chandra} and SKA.

\end{abstract}


%\linenumbers
%%%%%%%%%%%%%%%%%%%%%%%%%%%%%%%%%%%%%%
\section{Introduction}
%%%%%%%%%%%%%%%%%%%%%%%%%%%%%%%%%%%%%%

The high-energy universe is dominated by extreme and violently variable objects.
The powerful jets of relativistic plasma  associated with these sources are the largest and the most efficient particle accelerators known. 
The energy source,
the energy dissipation mechanism and the particle acceleration mechanism in these fast flares of non-thermal radiation remain puzzling. 
Proposed mechanisms include relativistic shocks and magnetic reconnection  
\citep{
2001ApJ...562L..63Z,
2002ApJ...578..763S,
2003ApJ...589..893L,2003ApJ...591..366K,
2004PhPl...11.1151J,
2005MNRAS.358..113L,2005ApJ...629..397P,
2006MNRAS.368.1561M,
2007ApJ...670..702Z,2007MNRAS.380...51K,
2009ARA&A..47..291Z,2009MNRAS.395L..29G,
2010ApJ...711...50T,
2011MNRAS.413..333N,2011MNRAS.418L..79T,
2012ApJ...754L..33C,2012PhRvL.108m5003H,
2014ApJ...783L..21S,0004-637X-806-2-167,2014PhRvL.113o5005G,
2015arXiv150802392N,2015arXiv151007243M,2015ApJ...809...97B,2015ApJ...804..111M,2015arXiv151204526T}.

One of the limitations to understanding these sources is our inability to localize the spatial origin of the emission. 
This failure is a direct result of the poor resolution of gamma-ray telescopes that reach, at best, an angular resolution of $0.1\,$deg. 
This angular resolution is unlikely to improve substantially with future instruments because it is limited by fundamental  physical effects including nuclear recoil. 

Gravitational lensing magnifies distant sources. 
Thus lensed gamma-ray blazars offer the
best opportunity for resolving the locations of the emitting regions. 
For example, the bright blazar PKS~1830-211 is a lensed system \citep{2011A&A...528L...3B}.
Analysis of  the time delays between gamma rays from the mirage images  of the  flaring episodes, combined with the lens model for this source
reveals that two gamma-ray flares originated from a region within 100~pc from the central engine.
Two additional gamma-ray flares originated at least $1.5\,$kpc from the central engine \citep{2015ApJ...809..100B}. 
The existence of  multiple variable emitting regions along the jet pose challenges for  understanding the particle acceleration mechanism. 
 
A second gravitationally lensed blazar B2~0218+35 offers further opportunities to explore the origin of the variable gamma-ray emission. 
This well-observed system has a number of features that enable the derivation of strong constraints on the nature of the gamma-ray source. 
The lens galaxy, observed with HST, is surprisingly simple and isolated \citep{2004MNRAS.349...14W}. 
There are extensive high-resolution radio observations of the lensed radio jet at several wavelengths. 
The Fermi-LAT light curve includes two flares, one of long duration and one short flare.

Based on the optical observations of the lens system and the radio observations of the lensed source, 
we demonstrate that the positions of the radio core and jet can be localized to 1 miliarcsecond.  
Localization of the gamma-ray source relative to the radio emitting regions requires both the time resolution of the Fermi-LAT light curve 
and a well-measured Hubble constant \citep{2014arXiv1403.5316B,2015ApJ...799...48B}.

The Fermi-LAT light curves provide time delays with an accuracy of a few hours. 
If the time delay originates from the resolved radio core, 
the associated Hubble constant should be the true value obtained with independent techniques. 
If there is an offset of even a few miliarcseconds between the gamma-ray emitting region and the radio core, 
the time delay will imply a Hubble constant that differs from the true value 
measured with many independent methods 
 \citep{2001ApJ...553...47F,2010ARA&A..48..673F,2010ApJ...711..201S,2011ApJ...730..119R,2011ApJ...732..129R,
2012ApJ...758...24F,2012MNRAS.425L..56C,2014MNRAS.440.1138E,2013ApJ...766...70S}.
Evidently, if the Hubble constant is well-known, the offset between the
gamma-ray emitting region and the radio core can be derived with remarkable significance 
limited only by the accuracy of the time delay and the independently determined Hubble constant \citep{2015ApJ...799...48B}.

We introduce the optical and radio observations of the gravitationally lensed blazar B2~0218+35 
and we reconstruct the properties of the lens and the source (Sections~\ref{sec:lens} and~\ref{sec:telescope}). 
In Section~\ref{sec:gamma}, we use the Fermi-LAT data to explore the gamma-ray properties of B2~0218+35. 
We focus on two gamma-ray flares where we  measure gravitationally induced time delays (Section~\ref{sec:TimeDelay}). 
In Section~\ref{sec:HPT}, we combine the measurements of the gamma-ray time delays, the well-resolved position of the radio core,
the reconstructed gravitational potential of the lens, and we explore the
relative spatial origin of the two gamma-ray flares. 
Finally, we apply the Hubble parameter tuning approach, 
where we use the independently measured Hubble constant to localize the gamma ray emission relative to the radio core.
We compare the results for B2 0218+35 with those for PKS~1830-211.
We discuss the implications of the source structure for gamma-ray emission mechanisms in Section~\ref{sec:Discussion}.
We also propose extension of this approach to sources that will be observed with SKA and {\it Chandra}.
We conclude in Section~\ref{sec:Conclusions}.

%%%%%%%%%%%%%%%%%%%%%%%%%%%%%%%%%%%%%%
\section{B2~0218+35: A Gravitationally-Lensed System }
\label{sec:lens}
%%%%%%%%%%%%%%%%%%%%%%%%%%%%%%%%%%%%%%

B2~0218+35 is a gravitationally-lensed system with the smallest known Einstein radius ($330\,$mas) \citep{1992AJ....104.1320O,1995MNRAS.274L...5P}. 
The system consists of a bright blazar  at redshift $z_S=0.944\pm 0.002$ \citep{2003ApJ...583...67C},
lensed by an apparently isolated spiral galaxy at redshift $z=0.6847$ \citep{1993MNRAS.263L..32B}.
The lens bends the emission of the jet into two bright images of the core and extended structures, including an Einstein ring 
\citep{1992MNRAS.254..655P,1992AJ....104.1320O,1993MNRAS.261..435P,1995MNRAS.274L...5P,2000MNRAS.311..389J,2001MNRAS.322..821B,2003MNRAS.338..599B}.

%The well-resolved radio images  show clear jet-like structures.
%\citet{2004MNRAS.349...14W} shows that the  jet sub-components 
%are located  exactly radially with respect to the center of mass of the lens. 
%The existence of the radio ring implies radial alignment of the jet on scales  $\lesssim$ kpc. 

The first measurement of the time delay using VLA $15\,$GHz polarization observations yielded a value of $12\pm3\,$days \citep{1996IAUS..173...37C}.
\citet{1999MNRAS.304..349B} used the results of a three-month VLA monitoring campaign at two frequencies 
and obtained  a time delay of $10.5 \pm 0.4\,$days. 
\citet{2000ApJ...545..578C} used high-precision VLA flux density measurements, over the same epoch as \citet{1999MNRAS.304..349B},
and measured a time delay of $10.1^{+1.5}_{-1.6}\,$days. 

This system has been a "golden lens" for Hubble constant measurement \citep{2004MNRAS.349...14W}.
However, despite precise measurements of the time delay, 
a clean lens environment without nearby companions or a surrounding cluster, 
and a negligible number of  structures along the line of sight which would complicate the modeling of the lens,
the H$_0$  values derived from this system are in the range 61-78~$\mbox{km\,s}^{-1}$Mpc$^{-1}$ 
\citep{2005MNRAS.357..124Y,2000ApJ...536..584L,2004MNRAS.349...14W}.
The most recent attempt to measure H$_0$ for B2~0218+35,  using a time delay of $11.46\pm0.16\,$days based on gamma-ray emission,
results in a Hubble constant of $64\pm4$~$\mbox{km\,s}^{-1}$Mpc$^{-1}$ \citep{HujCheung}.

This large scatter in the H$_0$ values can indicate complex source structure \citep{2015ApJ...799...48B}. 
To investigate the detailed source structure, we build a lens model.

%%%%%%%%%%%%%%%%%%%%%%%%%%%%%%%%%%%%%%
\section{B2~0218+35 as a High Resolution Cosmic Telescope}
\label{sec:telescope}
%%%%%%%%%%%%%%%%%%%%%%%%%%%%%%%%%%%%%%

\begin{figure}
%\vskip 1cm
\begin{center}
%Flare1_ACF.eps  UL not included in the CL
%/dane/Lensing/data/PKS1830/whole/mag_vs_distance_SIE.gp
\includegraphics[width=9.cm,angle=0]{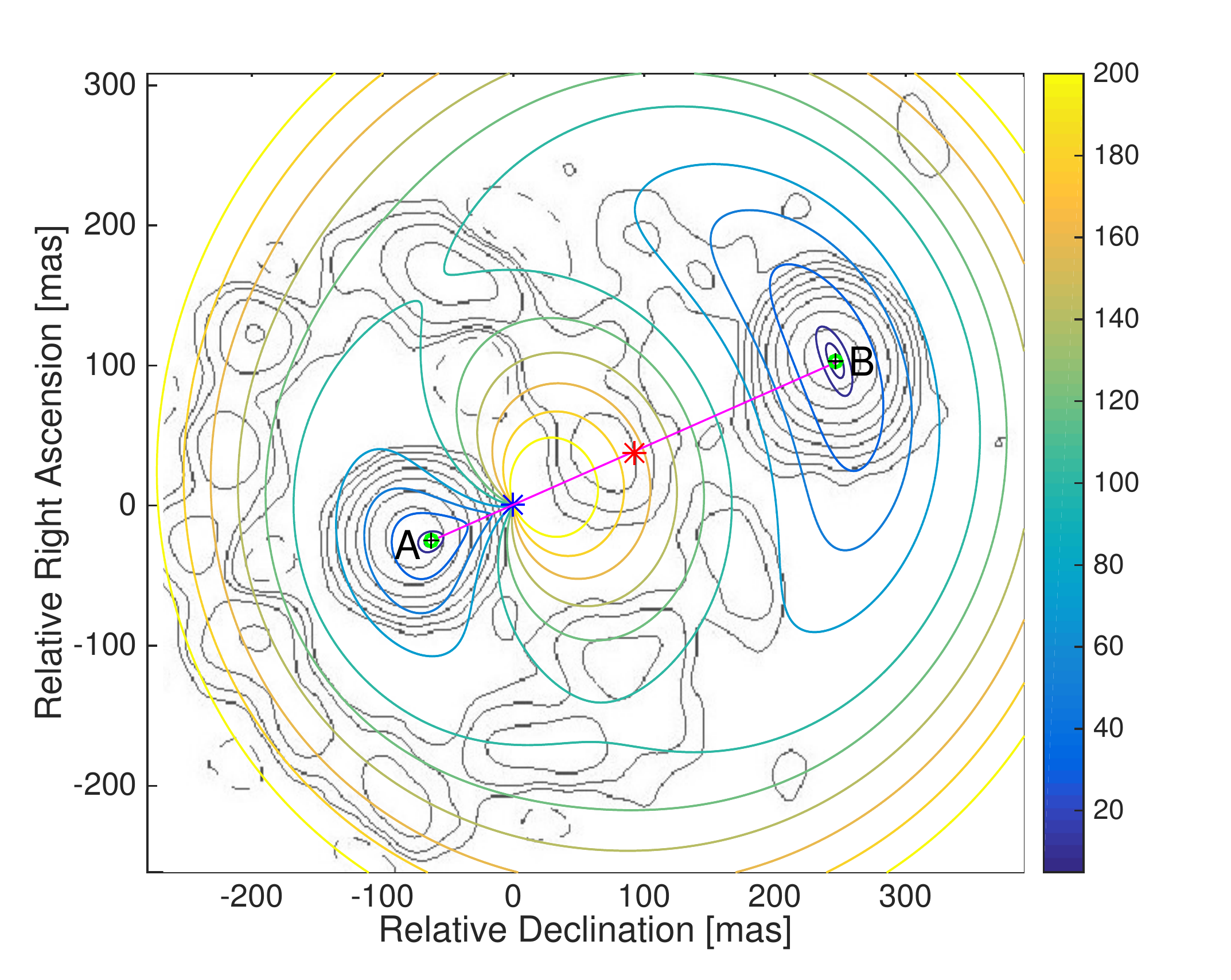}
\end{center}
\caption{\label{fig:Fermat} Image plane. The colored contours show the Fermat surface. 					 
					 The coordinates are relative to the reconstructed lens position.	
					  Black crosses show the reconstructed mirage image positions.
					  Green open circles show the positions of the 15~GHz mirage images of B2~0218+35.
					  The gray contours show the radio  emission observed  at 1.687~GHz. 
					  The magenta line indicates the axis connecting the mirage images. 
					  The blue star shows the  final  position of the lens center. 
					  The red star indicates the reconstructed position of the 15 GHz radio source.   
					        }
\end{figure} 

B2~0218+35 is a perfect system for lens modeling. 
The simplicity and  isolation of the lensing galaxy results in a clean gravitational lens potential. 
Previous lens models show that the observations are consistent with a Singular Isothermal Sphere (SIS) model for the mass distribution of the lens \citep{2004MNRAS.349...14W,2011AstL...37..233L}.
These studies focused on the properties of the lensing galaxy and  measurement of the Hubble parameter \citep{2005MNRAS.357..124Y}.
Here, we  use the lens as a high-resolution telescope to investigate the structure of the source over wavelengths ranging from radio to gamma-ray energies. 
We first use the lens model to determine an accurate position for the radio core. We evaluate the uncertainties in the model using Monte Carlo simulations.
%

%%%%%%%%%%%%%%%%%%%%%%%%%%%%%%%%%%%%%%
\subsection{Constraining the Lens}
\label{sec:conlens}
%%%%%%%%%%%%%%%%%%%%%%%%%%%%%%%%%%%%%%

We base the model on radio VLBA observations at 15~GHz, where the position of 
mirage images of the core are measured with 0.6~mas accuracy \citep[1994 Oct 3,][]{1995MNRAS.274L...5P}.
The position of the mirage image B (brighter image located outside the Einstein ring;
green circle in Figure~\ref{fig:Fermat}) with  respect to  image A (green circle, Figure~\ref{fig:Fermat}), 
along with all the parameters, are  summarized in Table~\ref{tab:InputParameters}.

%---------------------- Table: Input Parameters -------------------------------%
\begin{table}[h!]
  \begin{center}
    \caption{Input Parameters for the Lens Modeling}
    \label{tab:InputParameters}
    \begin{tabular}{lc}
   \hline
      Parameter &  Value\\
   \hline
      Image A  at 15 GHz & (0,0) \\
      Image B  at 15 GHz &   $(309.2,-127.4)\pm0.6\,$mas \footnote{\citet{1995MNRAS.274L...5P}} \\
      Estimated Source Position & $(154.6,-63.7)\pm10\,$mas \\
      Einstein Angle & $\theta_E = 167.2\pm 0.6\,$mas \\
       Approximate Lens Position & $(56,-23)\pm20\,$mas  \\
      Optical Center of the Galaxy \footnote{\citet{2005MNRAS.357..124Y}}& \\
     -- no masking     & $(57\pm4,1\pm6)$ \\
      -- spiral arms masked    & $(75\pm6,-6\pm13)$\\
      Position Angle& $\phi_0=49^\circ$\footnote{\citet{2005MNRAS.357..124Y}} \\
      Time Delay at 15 GHz & $10.1\pm1.6\,$days\footnote{\citet{2000ApJ...545..578C} }, \\
      					&  $10.5\pm0.4\,$days \footnote{\citet{1999MNRAS.304..349B}}\\
      Magnification Ratio at 15 GHz & $3.623\pm0.065$\footnote{\citet[][]{1995MNRAS.274L...5P}} \\

      Source Redshift & $z_S=0.944$\footnote{\citet{2003ApJ...583...67C}}, \\
      Lens (Galaxy) Redshift &  $z_L=0.6847$\footnote{\citet{1993MNRAS.263L..32B}} \\
   \hline
    \end{tabular}
  \end{center}
\end{table}
%--------------------------------------------------------------------------------------%

The positions of the 15 GHz mirage images yield a measure of the Einstein radius and the position of the source. 
The Einstein radius of the lens, with a mass distribution close to a SIS, 
is half the distance between the mirage images; $\theta_E = (\theta_A+\theta_B)/2 = 167.2\pm 0.6\,$mas. 
The  corresponding lens mass within one Einstein radius is  $\sim 2\times 10^{10}\,M_{\odot}$. 
For a lens of such a small mass, resolution of the mirage images is possible only with
high-resolution imaging.

The mirage images appear on the axis defined by the position of the source and the lens
(magenta line in Figure~\ref{fig:Fermat}), 
at  distances  of $\pm \theta_E$ from the source. 
The source is located at  half the distance between the mirage images, $\theta_S = (\theta_A-\theta_B)/2$. 
The source position may deviate  from the axis if the mass distribution deviates from a SIS. 
In the model described below, we thus search for the best source position within $10\,$mas from this estimate.  

The last, major unknown is the position of the lens. 
\citet{2005MNRAS.357..124Y} derived the optical center of the galaxy with an accuracy of $\sim15\,$mas (see Table~\ref{tab:InputParameters}). 
We seek a center of mass reconstructed  to $\sim1\,$mas.
We first take the position of the lens as inferred from the optical images.
Then, as an additional constraint, we use the lens geometry;
the lens must be located close to the image axes and to the center of the Einstein ring 
(Figure~\ref{fig:Fermat}). 

For demonstration purposes, in Figure~\ref{fig:Fermat} we display the contours of radio  emission observed  at 1.687~GHz 
with a beam FWHM of $50\times50\,$mas.
However, this image was available to us only in jpg format  downloaded from JVAS
\footnote{http://www.jb.man.ac.uk/research/gravlens/lensarch/ B0218+357/B0218+357.html}. 
We rescaled it by $\sim5$\% to align it with well-resolved mirage images at 15~GHz. 
The image at 1.687~GHz resolves the mirage images of the core and the Einstein ring structures.
After rescaling,  the center of the ring indicates a plausible position of the lens with an accuracy of $\sim50\,$mas.
Using all of this information, we find the approximate position of the lens (see Table~\ref{tab:InputParameters}).
In the lens model described below, we use Monte Carlo simulations to refine the best lens position within this region. 

The VLBA observations at 15~GHz also provide  
a very precise flux density ratio between mirage images of $3.623\pm 0.065$. 
However, observations of B2~0218+35 at different frequencies and epochs 
show a large spread in measured magnification ratios, from 1 to 6 \citep{1999MNRAS.304..349B,2000ApJ...545..578C,HujCheung}.
This large spread in magnification ratio may  result from different propagation effects \citep{2007A&A...465..405M}, 
substructures in the mass distribution \citep{1998MNRAS.295..587M,2004ApJ...610...69K,2012MNRAS.419.3414M,2004MNRAS.350..949B}, 
or even from microlensing \citep{2015arXiv150701092V}. 
To avoid this additional complexity, we do not use the flux ratio as a constraint. 
However, the best-fit model does yield a magnification ratio that is consistent with the range of values observed at other frequencies

%%%%%%%%%%%%%%%%%%%%%%%%%%%%%%%%%%%%%%
\subsection{Lens Modeling}
%%%%%%%%%%%%%%%%%%%%%%%%%%%%%%%%%%%%%%

We investigate the properties of the lens system with a  {\tt MATLAB} code inspired by \citet{2009MNRAS.396.1985Z,2013ApJ...762L..30Z}. 
We construct a  $800\times 800\,$mas grid with a resolution of $1\,$mas. We define 
coordinates relative to the position of  the lens center. 
We perform all calculations in the image plane because the observed positions of the images are directly   linked to the image plane, not to the source plane. 

%%%%%%%%%%%%%%%%%%%%%%%%%%%%%%%%%%%%%%
\subsubsection{Gravitational Lensing Formalism}
%%%%%%%%%%%%%%%%%%%%%%%%%%%%%%%%%%%%%%

We compute the Fermat surface using Eq.~(61) from \citet{1996astro.ph..6001N}:
\begin{equation}
(\vec{\theta} - \vec{\beta}) - \vec{\nabla}_\theta \psi = 0 \,,
\end{equation}
where $\vec{\theta}$ is the position of a mirage image, $\vec{\beta}$ is the source position, and $\psi$ is the gravitational potential of the lens. 
The Fermat principle implies that the images form at the extrema (maxima, minima, and saddle points) of the surface \citep{1986ApJ...310..568B}. 
We then  search for the extrema of the Fermat surface using {\tt Matlab} procedure {\tt extrema2}\footnote{http://www.mathworks.com/matlabcentral/fileexchange/12275-extrema-m--extrema2-m}. 
Figure~\ref{tab:Fit} shows color contours of the Fermat surface with two minima where the mirage images form. 
We find the positions of these minima and calculate the offset between their coordinates 
and positions of the mirage images resolved at 15~GHz.

The position of the images and the gravitational potential of the lens allow us to calculate 
the time delay between the mirage images.
We use Eq~(63) from \citet{1996astro.ph..6001N}:
\begin{equation}
\label{eq:dt}
t(\vec{\theta}) = D\frac{(1+z_L)}{c} \bigg[\frac{1}{2}(\vec{\theta} - \vec{\beta}) ^2-\psi(\vec{\theta})\bigg]\,,
\end{equation}
to calculate the time delay between the arrival of photons from image $\theta_A$ and $\theta_B$. 
We then calculate the difference between the estimated  time delay and the time delay measured at 15~GHz. 

We calculate magnifications of the mirage images using Eqs~(55-60) from \citet{1986ApJ...310..568B}.
This calculation is also based on the gravitational potential and the positions of the mirage image. 
We do not, however,  use the magnification ratio to extract the lens parameters.
We provide the formalism because we do use the magnification later to constrain the origin of the gamma-ray radiation.

%%%%%%%%%%%%%%%%%%%%%%%%%%%%%%%%%%%%%%
\subsubsection{Finding a Unique Mass Model}
%%%%%%%%%%%%%%%%%%%%%%%%%%%%%%%%%%%%%%

We seek the gravitational potential along with source and  lens locations that reproduce the observations.
We compare the reconstructed positions of the lensed images with well-resolved mirage images at 15~GHz. 
As an additional constrain, we use time delay measured at 15~GHz.  
%The resolved images alone cannot provide unique mass distribution of the lens.
%In principle, there can be a range of solutions that could reconstruct the position of the images. 

To find the lens solution, we repeat  our calculations of the image positions and the time delay between them for a range of parameters. 
We investigated a range of complex models for the gravitational potential using the Monte Carlo simulations. 
We added parameters including  a core, a variable slope for the mass distribution, 
and a variable ellipticity and position angle of the lensing galaxy.
None of the added parameter to the lens model where able to improve the fit and reconstruct the observations with desired accuracy.
We vary the lens position around the value listed in Table~\ref{tab:InputParameters}. 
We explore a region of  $20\,$mas with a $1\,$mas step.  
We search for the best source position around  the value listed in Table~\ref{tab:InputParameters}. 
{\bf In Appendix~\ref{app},  we describe our Monte Carlo simulations and our investigation of systematic errors associated with lens model.}

We define the best-reconstructed model as the one which reproduces the positions of the mirage images with the smallest offset
and where the time delay is within $1\,\sigma$ of the measured time delay at 15~GHz.  

%%%%%%%%%%%%%%%%%%%%%%%%%%%%%%%%%%%%%%
\subsubsection{The Best Model}
%%%%%%%%%%%%%%%%%%%%%%%%%%%%%%%%%%%%%%

We achieved the best reconstruction for an elliptical singular isothermal sphere \citep{KneibManual}:
\begin{equation}
\psi(r,\theta) = r \theta_E  \sqrt{1-\epsilon \cos (2(\phi - \phi_0))} \,,
\end{equation}
where $\epsilon$ is an ellipticity of the gravitational potential,
$\phi_0$ is the position angle of the potential,
and $\theta_E$ is an Einstein angle defined as:  
\begin{equation}
\label{eq:Potential}
 \theta_E = 4 \pi \frac{\sigma^2_0}{c^2} \frac{D_{LS}}{D_{OS}} \,,
\end{equation}
where $\sigma_0$ is the central velocity dispersion of the 3D velocity field, 
and $D_{LS}$ and $D_{OS}$ are  cosmological distances from the lens to the source,
and from the observer to the source, respectively.
We also define:
\begin{equation}
 D\equiv \frac{D_{OL}D_{OS} }{D_{LS}}=hd\,,
 \end{equation}
where $D_{OL}$ is the distance from the observer to the lens.
The parameter $h$ refers to the Hubble constant, H$_0=h\times$~100$\,\mbox{km\,s}^{-1}$Mpc$^{-1}$. 
We calculate distances based on a homogenous Friedmann-Lema{\^i}tre-Robertson-Walker cosmology,
using $h=0.673$, the mean mass density $\Omega_M=0.315$ and the normalized 
cosmological constant $\Omega_\Lambda=0.686$ \citep{2013arXiv1303.5076P}.

%%%%%%%%%%%%%%%%%%%%%%%%%%%%%%%%%%%%%%
\subsubsection{Error Estimation}
%%%%%%%%%%%%%%%%%%%%%%%%%%%%%%%%%%%%%%

To estimate statistical errors, we use Monte Carlo chain simulations.
We based our algorithm on the MCMC toolbox for {\tt Matlab}\footnote{http://helios.fmi.fi/~lainema/mcmc/} \citep{MCMC}.

We test for systematics in our simulations by comparing the numerical solution with an analytic model. 
The simplest analytic solution is the SIS. We compare  the positions of the images, time delays and magnification ratios for the SIS
calculated analytically and numerically for positions of the sources across the entire lens plane. The numerical procedure applied on a grid with a  $1\,$mas resolution,
on average reconstructs image positions with $\sim0.3\,$mas. 
On average, the time delay is reproduced within $0.01\,$days,
and the magnification ratio within 0.05.  
This level of precision shows that on a scale of $1\,$mas,
our lens parameters are unaffected by systematic numerical errors. 

%---------------------- Table: Fit -------------------------------%

\begin{table}[h!]
  \begin{center}
    \caption{Results of the Fit}
    \label{tab:Fit}
    \begin{tabular}{ccc}
   \hline
      Parameter &  Value\\
   \hline
      Ellipticity $\epsilon$ & $0.0057\pm0.0042$ \\
      Source Position $(x_S,y_S)$ & $(154.2\pm0.8,-62.9\pm0.7)\,$mas \\
      Lens Position $(x_L,y_L)$ & $(62.2\pm0.9,-25.0\pm0.8)\,$mas \\      
   \hline
    \end{tabular}
  \end{center}
\end{table}

%--------------------------------------------------------------------%

%%%%%%%%%%%%%%%%%%%%%%%%%%%%%%%%%%%%%%
\subsubsection{Lens Modeling Results}
\label{sec:LensResults}
%%%%%%%%%%%%%%%%%%%%%%%%%%%%%%%%%%%%%%

Table~\ref{tab:InputParameters} summarizes the input parameters for the lens model;
the ellipticity of the lens, and the source and lens positions. 
Table~\ref{tab:Fit} shows the model results along with the statistical errors from
the Monte Carlo chain simulations. 

The fit yields an $\epsilon \sim 0$,  essentially an  isotropic SIS. 
We reconstruct the lens and source positions with an accuracy of $1\,$mas
corresponding to $8\,$pc in the source plane. 

The positions of the mirage images are most sensitive to changes of the source position in the tangential direction relative 
to the images-lens axis. 
Changing the source position by $1\,$mas in the tangential direction,
moves the  mirage images by $5.5\,$mas. 
In the radial direction a $1\,$mas change in the source position  displaces the image  by only $2.7\,$mas. 
Table~\ref{tab:Reconstruction}  and Figure~\ref{fig:Fermat}  
show that the model reproduces the observed mirage image positions to $0.4-0.8\,$mas.

%---------------------- Table: Reconstructed Parameters -------------------------------%
\begin{table}[h!]
  \begin{center}
    \caption{Reconstruction}
    \label{tab:Reconstruction}
    \begin{tabular}{ccc}
   \hline
      Parameter &  Value & Difference\\
   \hline
       Image A & (-0.4,0) & $0.4\,$mas \\
       Image B &  $(308.6,-128.0)$ & $0.85\,$mas  \\
      Time Delay & $10.7\,$days & $\sim0.2\,$days\\
      Magnification Ratio & $3.85$ & 0.23 \\
   \hline
    \end{tabular}
  \end{center}
\end{table}

\begin{figure*}
%\vskip 1cm
\begin{center}
%Flare1_ACF.eps  UL not included in the CL
%/dane/Lensing/data/PKS1830/whole/mag_vs_distance_SIE.gp
\includegraphics[width=5.cm,angle=-90]{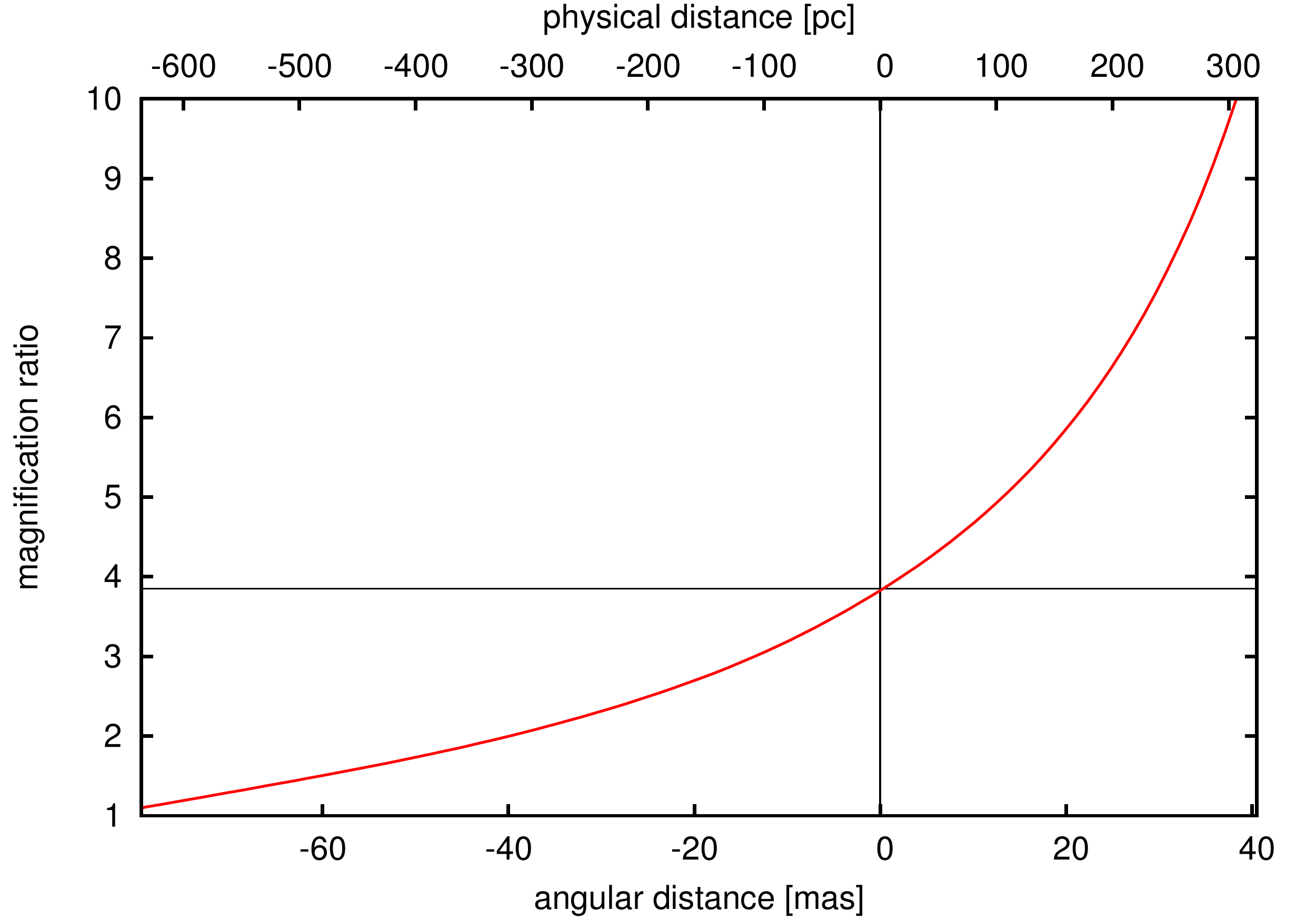}
\includegraphics[width=5.cm,angle=-90]{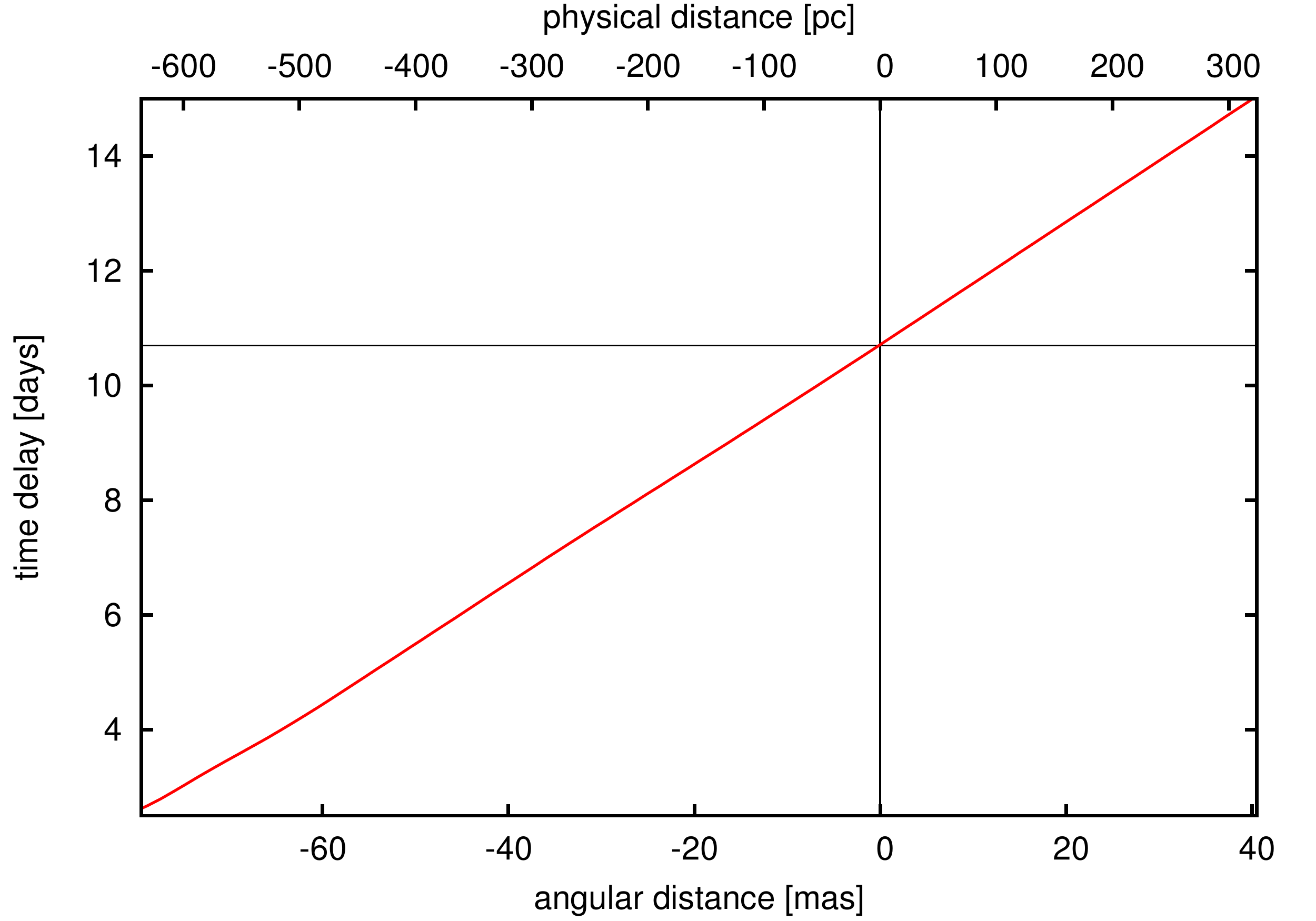}
\end{center}
\caption{\label{fig:jet_lens} Magnification ratios  ({\bf Left}) and time delays ({\bf Right}) as a function of the distance between the core and  emitting regions along the jet.
					   {\bf Left}: 
					             	 The gray lines indicate the magnification ratio, 3.85,  
						           expected at the position of  the 15~GHz core. 
					   {\bf Right}:
					   		 The gray lines indicate the time delay of 10.7 days
						           expected for an emitting region coincident with the position of the 15~GHz core. 
	       }
\end{figure*} 

%%%%%%%%%%%%%%%%%%%%%%%%%%%%%%%%%%%%%%
\subsection{Lens Model and the Jet Alignment}
\label{sec:LensJet}
%%%%%%%%%%%%%%%%%%%%%%%%%%%%%%%%%%%%%%

We investigate the origin of variable emission along the relativistic  jet. 
The alignment of the jet and the mass distribution of the lens are necessary to predict the range of time delays and corresponding magnification ratios. 

The alignment of the jet of B2~0218+35 is known from the well-resolved radio images which  show clear jet-like structures.
\citet{2004MNRAS.349...14W} show that the  jet sub-components are oriented exactly radially with respect to the center of mass of the lens. 
The existence of the radio Einstein ring implies radial alignment of the jet on scales  $\lesssim$ kpc. 

We use the alignment of the jet and the lens model (Table~\ref{tab:Fit}) to calculate the time delay and corresponding magnification ratio along the jet.
Figure~\ref{fig:jet_lens} shows the result.

The time delay is very sensitive to the distance of the source from the center of the lens. 
The radial alignment of the jet of B2~0218+35 produces maximal differences in the time delay among regions 
distributed along the jet. 
In the radial direction the time delay changes by $0.13\,$days ($3.12\,$hours) per $1\,$mas
but in the tangential direction, it changes much more slowly, $0.01\,$day per $1\,$mas.

%%%%%%%%%%%%%%%%%%%%%%%%%%%%%%%%%%%%%%
\section{B2~0218+35 as a Gamma-Ray Source}
\label{sec:gamma}
%%%%%%%%%%%%%%%%%%%%%%%%%%%%%%%%%%%%%%
%-------------------------------- Count Map --------------------------------%
%/dane/Lensing/data/B2_0218/whole/B20218_cmap.eps
\begin{figure}
%\vskip 1cm
\begin{center}
\includegraphics[width=8.5cm,angle=0]{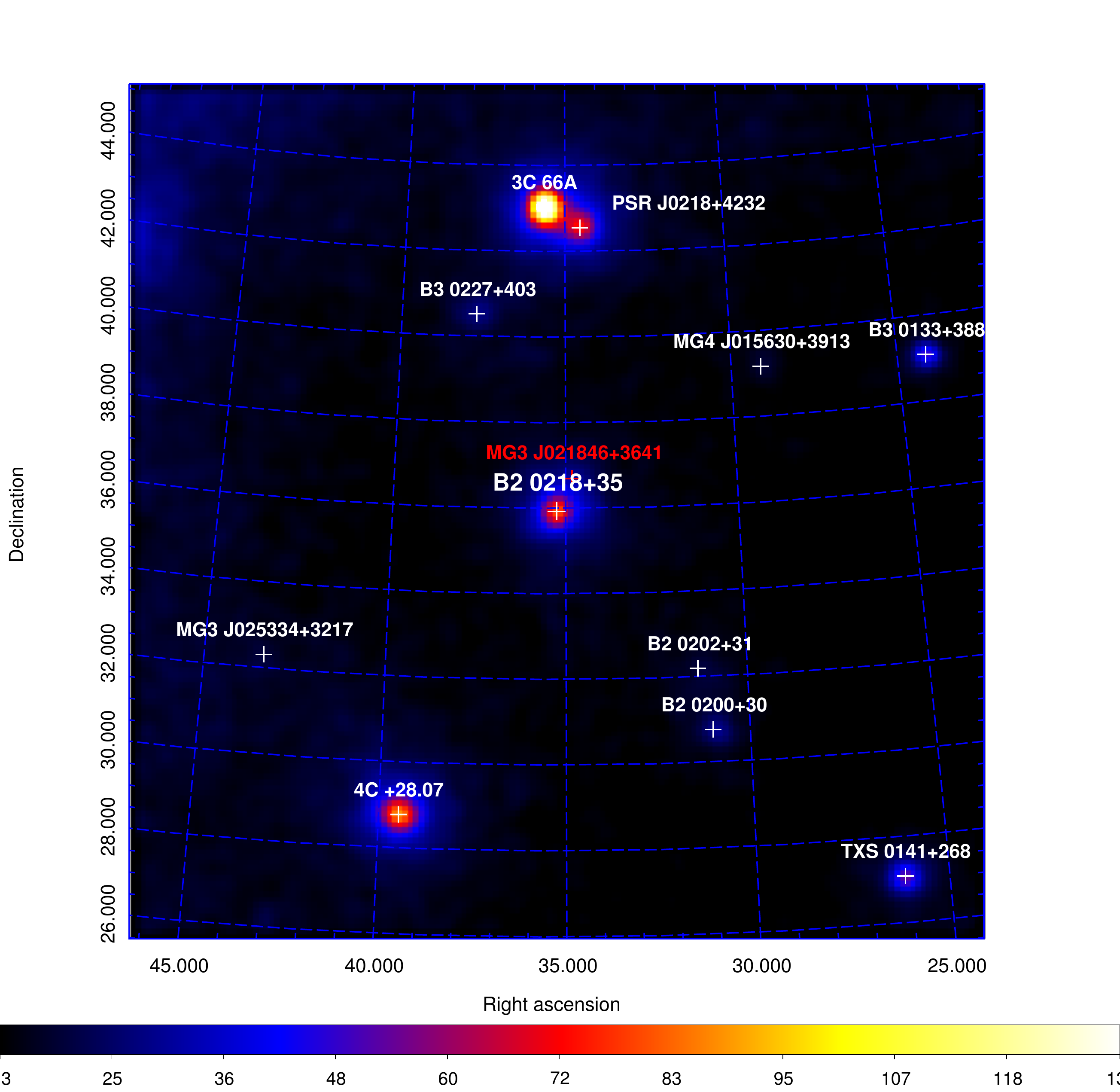}
\end{center}
\caption{\label{fig:counts_map} 
                           {\it Fermi}-LAT count map of B2~0218+35.
                           The energy range is 100 MeV to 300 GeV. }
\end{figure}

%-------------------------------- Light Curves --------------------------------%

%-------------------------------- Whole--------------------------------%
%/dane/Lensing/data/B2_0218/whole/d7_B20218_whole.pdf
\begin{figure*}
%\vskip 1cm
\begin{center}
\includegraphics[width=17.5cm,angle=0]{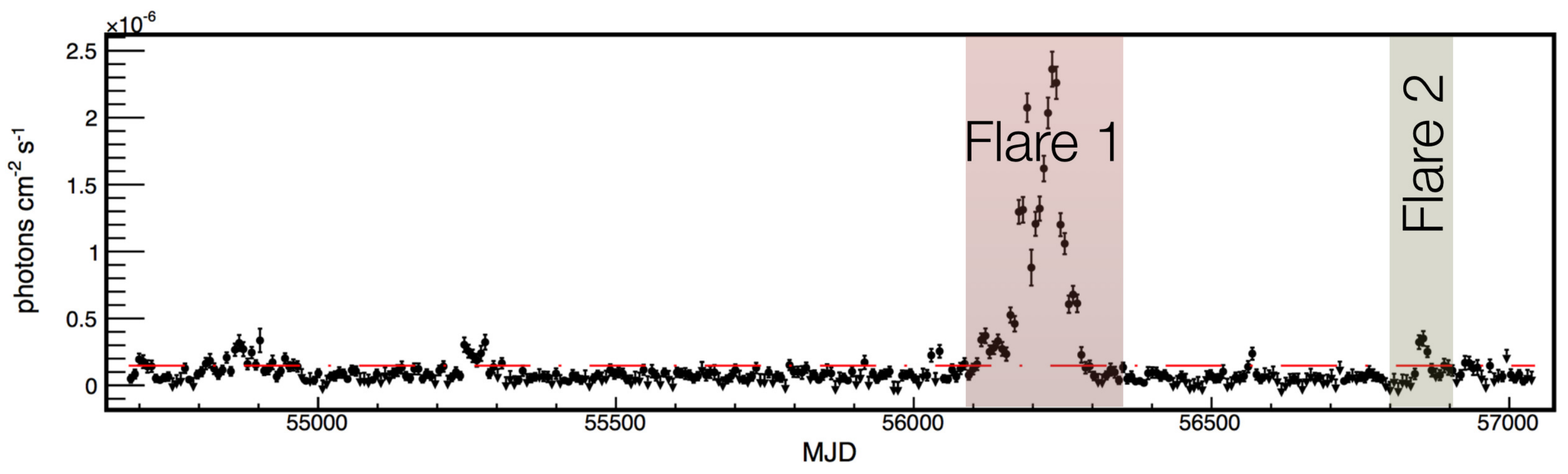}
\end{center}
\caption{\label{fig:lc_whole} 
                           {\it Fermi}-LAT light curve of B2~0218+35 with seven-day binning.
                           The energy range is 100 MeV to 300 GeV. }
\end{figure*}

\begin{figure*}
%\vskip 1cm
\begin{center}
%-------------------------------- Flare 1--------------------------------%
%/dane/Lensing/data/PKS1830/flare1/d1_d05_B20218_flare1.pdf
%\includegraphics[width=18.5cm,angle=0]{plots/B20218_Flare1_lc.pdf} \\
%\includegraphics[width=18.5cm,angle=0]{plots/B20218_Flare2_lc.pdf} \\
\includegraphics[width=8.7cm,angle=0]{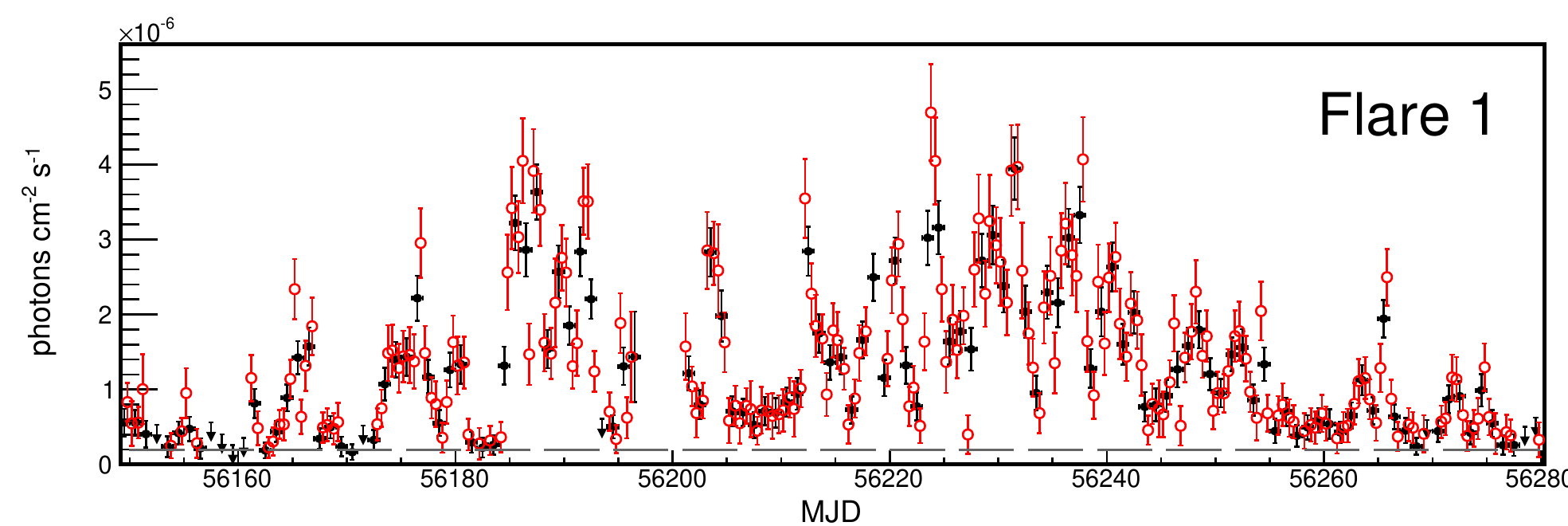} 
%-------------------------------- Flare 2--------------------------------%
%/dane/Lensing/data/B2_0218/flare2/d4_d2_B20218_Flare2.pdf
\includegraphics[width=8.7cm,angle=0]{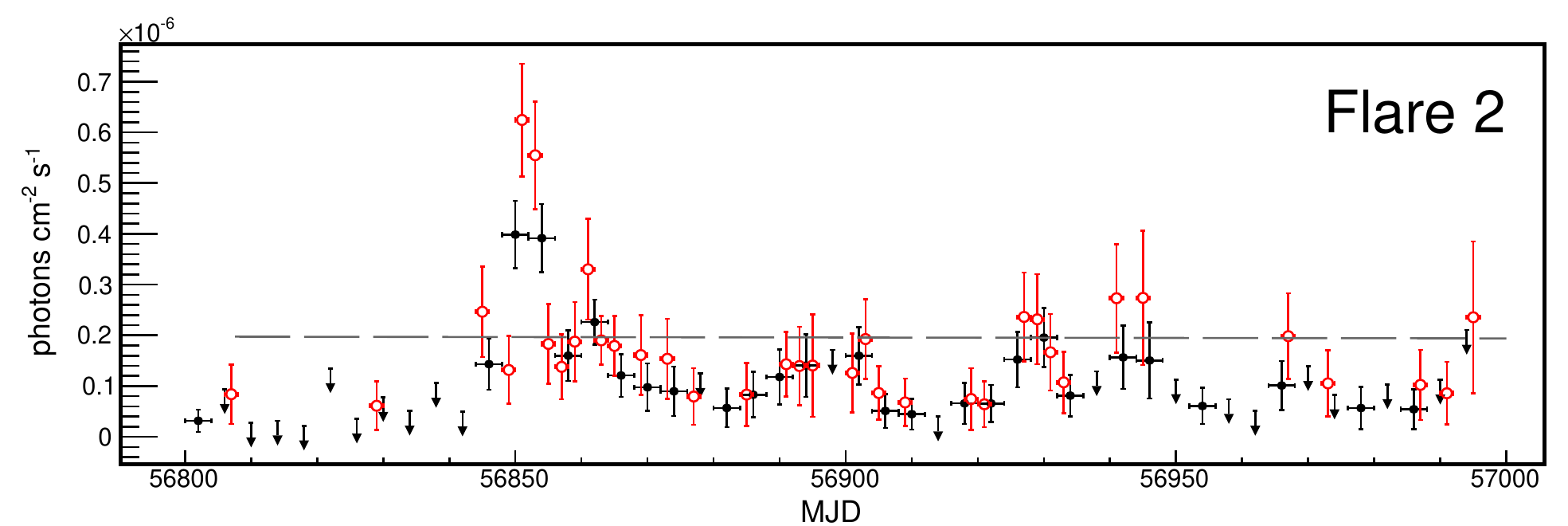} \\
\end{center}
\caption{\label{fig:lc_flares} 
                           {\it Fermi}-LAT light curves of B2~0218+35 during two flaring periods. 
                             The green dashed line represents the average flux ($1.41\pm0.04 \times 10^{-7}\, \mbox{photons}\, \mbox{cm}^{-2}\,\mbox{s}^{-1}$) 
                           measured from the 6.5 year light curve of B2~0218+35 in the energy range  from 100 MeV to 300 GeV. 
                           {\bf Top:} %The light curve of the first flaring period 
                           			Flare~1 with 12-hour binning (black filled circles)
                                		and with 6-hour binning (red open circles).
                           {\bf Bottom:} %The light curve around the second active period
                           Flare~2 with  four-day binning (black filled circles),
                           and  two-day binning (red open circles).
}
\end{figure*}

 B2~0218+35 is a bright gamma-ray source. 
Here, we analyze the gamma-ray light curve observed with 
the  Large Area Telescope onboard the {\it Fermi} mission \citep[{\it Fermi}-LAT,][]{2009ApJ...697.1071A}. 

We analyze the {\it Fermi}-LAT {\tt P7REP} events and spacecraft data of B2~0218+35 
during the period MJD: 54682--57041. 
We use the standard likelihood tools distributed with 
the {\tt Science Tools v9r32p5} package available on the {\it Fermi} Science Support Center webpage.

We only used  events in the {\tt CLEAN} dataset with the highest probability of being photons. 
We exclude events with zenith angles $>$~100$^\circ$  to 
limit contamination by Earth albedo gamma rays
produced by cosmic rays interacting with the upper atmosphere.
We  also remove events with rocking angles $>$~52$^\circ$ 
to eliminate time intervals  when the Earth entered the LAT Field of View (FoV).

The selected events with reconstructed energies above 100~MeV within a square region of 10$^\circ$ radius 
are centered on the coordinates of B2~0218+35 (see Figure~\ref{fig:counts_map}).
We analyze the selected photons  with a binned maximum likelihood method \citep{1996ApJ...461..396M}. 

We model the background emission  using a galactic diffuse emission model ({\tt gll\_iem\_v05}) 
with an isotropic component ({\tt iso\_clean\_v05}; available on the {\it Fermi} Science Support Center webpage). 
The fluxes are based on the post-launch instrument response function {\tt P7REP\_CLEAN\_V15}. 

The {\tt XML} source model contains all of the sources  included in the Second {\it Fermi}/LAT catalog \citep{2012ApJS..199...31N} 
within a radius  of 20$^\circ$ around B2~0218+35.
We first analyze the {\tt XML} source model fitting the sources within 10$^\circ$;
within an annulus from 10$^\circ$ to 20$^\circ$ we fix the sources to their 2FGL values.
We calculate the test statistic (TS)  for all the sources  located within the 10$^\circ$ radius 
during the time period MJD: 54682--57041.
Sources with a TS lower than 6.5, 
corresponding to a statistical significance of $\approx 2.5\, \sigma$,
are then fixed in further analysis. 
The resulting {\tt XML} source model is then the basis for the 6.5 year light curve in Figure~\ref{fig:lc_whole}. 
The TS is also based on this source model.

The TS of B2~0218+35 for this dataset is 8890,
corresponding to a statistical significance of $\approx 94\, \sigma$.
The energy spectrum is best described by a power law with 
$\Gamma = 2.28 \pm 0.02$ 
and an integral flux of 
$F(0.1-300\,\mbox{GeV})=(1.42\pm0.05)\times10^{-7}\,\mbox{ph}\,\mbox{cm}^{-2}\mbox{s}^{-1}$.
The highest energy event recorded by {\it Fermi}/LAT corresponds to $\sim 95.7\,$GeV.
The flux in each time bin is reconstructed in the same manner as
for the full time range; the photon index and integral flux in these bands are free parameters. 

Our goal is to investigate the spatial origin of gamma-ray flares.
Thus, we look for periods of flaring activity in the gamma-ray light curve of B2~0218+35.
We define a flare as a period of time when the emission in a one-day bin 
increases by at least two sigma relative to the average flux. 
Note that Figure~\ref{fig:lc_whole}  shows data in 7-day bins. 

Based on our definition, we identify two active periods;
a very long flare  between MJD: 56160 -- 56280, 
and a short flare consisting of a single bright event, 
occurring between  MJD: 56800 -- 57000. 
Figure~\ref{fig:lc_flares} shows both flares. 
Visually it may seem that there are other flares in Figure~\ref{fig:lc_whole}. 
However, these apparent flares are not significant with one-day bins. 
We use  the light curves for the two significant flares to compute time delays.

%%%%%%%%%%%%%%%%%%%%%%%%%%%%%%%%%%%%%%
\section{Time Delay Measurement}
\label{sec:TimeDelay}
%%%%%%%%%%%%%%%%%%%%%%%%%%%%%%%%%%%%%%

In the B2~0218+35 system, radio observations reveal two images separated by $\sim330\,$mas.
The angular resolution of the {\it Fermi}-LAT detector is $\sim1\,$deg.
Thus the observed gamma-ray light curve is  a sum of two unresolved mirage images shifted in time,
but with a constant magnification ratio. 
The temporal resolution of the {\it Fermi}-LAT detector  allows determination of the time delay at gamma ray energies. 
With  sufficiently well-determined time delays, the relationship between the radio and gamma-ray time delays constrain the relative source positions.

The {\it Fermi}-LAT data cover a long, uninterrupted periods containing the two significant flares. 
There are several methods for extracting the time delay from these binned data \citep{2015ApJ...809..100B}.
We use the standard Autocorrelation Function (ACF), 
and  the more sensitive Double Power Spectrum (DPS) method \citep{2011A&A...528L...3B,2013arXiv1307.4050B}.
The detailed description of the DPS method together with comparison to other methods is described in \citet[Section~3.2.2 and Appendix A,][]{2015ApJ...809..100B}
The DPS is similar to the Cepstrum method \citep{Bogert1963},
where a time series with a delay transforms  into Fourier space with the extra component $e^{-2\pi i f a}$,
where $f$ is frequency  and $a$ is the delay. 
Squaring the absolute value of this extra component results in a periodic pattern imprinted on the power spectrum.
The period of this pattern in the frequency domain  is the inverse of the relative time delay $a$.
We identify the time delay by calculating and analyzing the power spectrum of the power spectrum that includes the periodic pattern. 
We apply these methods to the first flaring period in Section~\ref{sec:dt_Flare1}.

Monte Carlo simulations demonstrate that this signal processing allows removal of the intrinsic variability of the source. 
We have previously demonstrated that this procedure
yields precise and significant estimates of the time delays \citep{2015ApJ...809..100B}.   

The DPS method is very efficient when applied to long, evenly sampled light curves.
However, sometimes the flare is an isolated event  of very short duration like Flare 2.
To treat these flares, we use the Maximum Peak Method  \citep[MPM, Section 3.2.3,][]{2015ApJ...809..100B} 
where  we  calculate the ratio between the flux in the flare and flux in the subsequent data. 
We compare the flux ratios as a function of the lag between the brightest  flare 
and the subsequent light curve with the predictions of the model (Section~\ref{sec:dt_Flare2}) of B2~0218+35.

%%%%%%%%%%%%%%%%%%%%%%%%%%%%%%%%%%%%%%
\subsection{Flare 1}
\label{sec:dt_Flare1}
%%%%%%%%%%%%%%%%%%%%%%%%%%%%%%%%%%%%%%

\begin{figure*}
%\vskip 1cm
\begin{center}
\includegraphics[width=5.5cm,angle=-90]{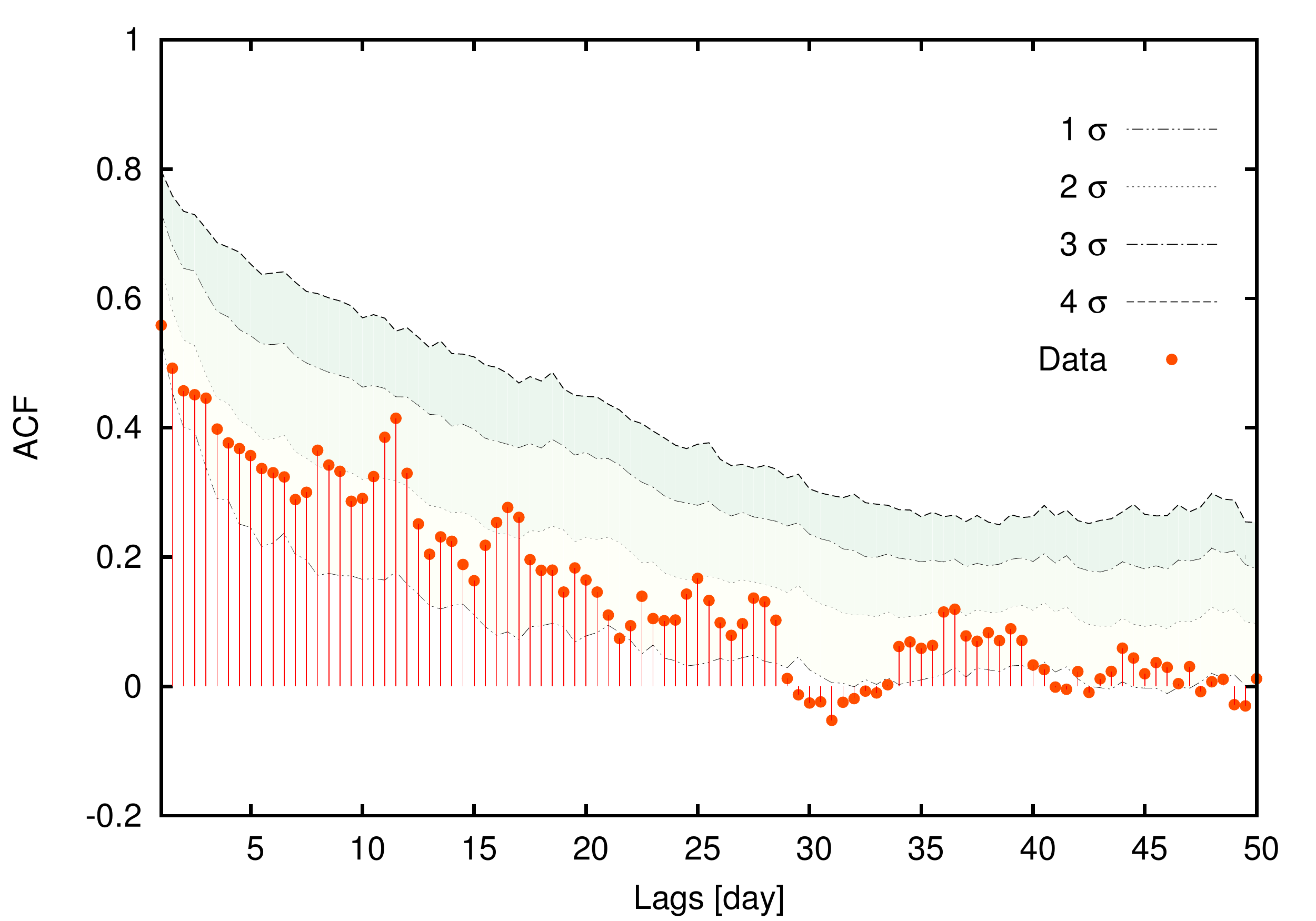}
\includegraphics[width=5.5cm,angle=-90]{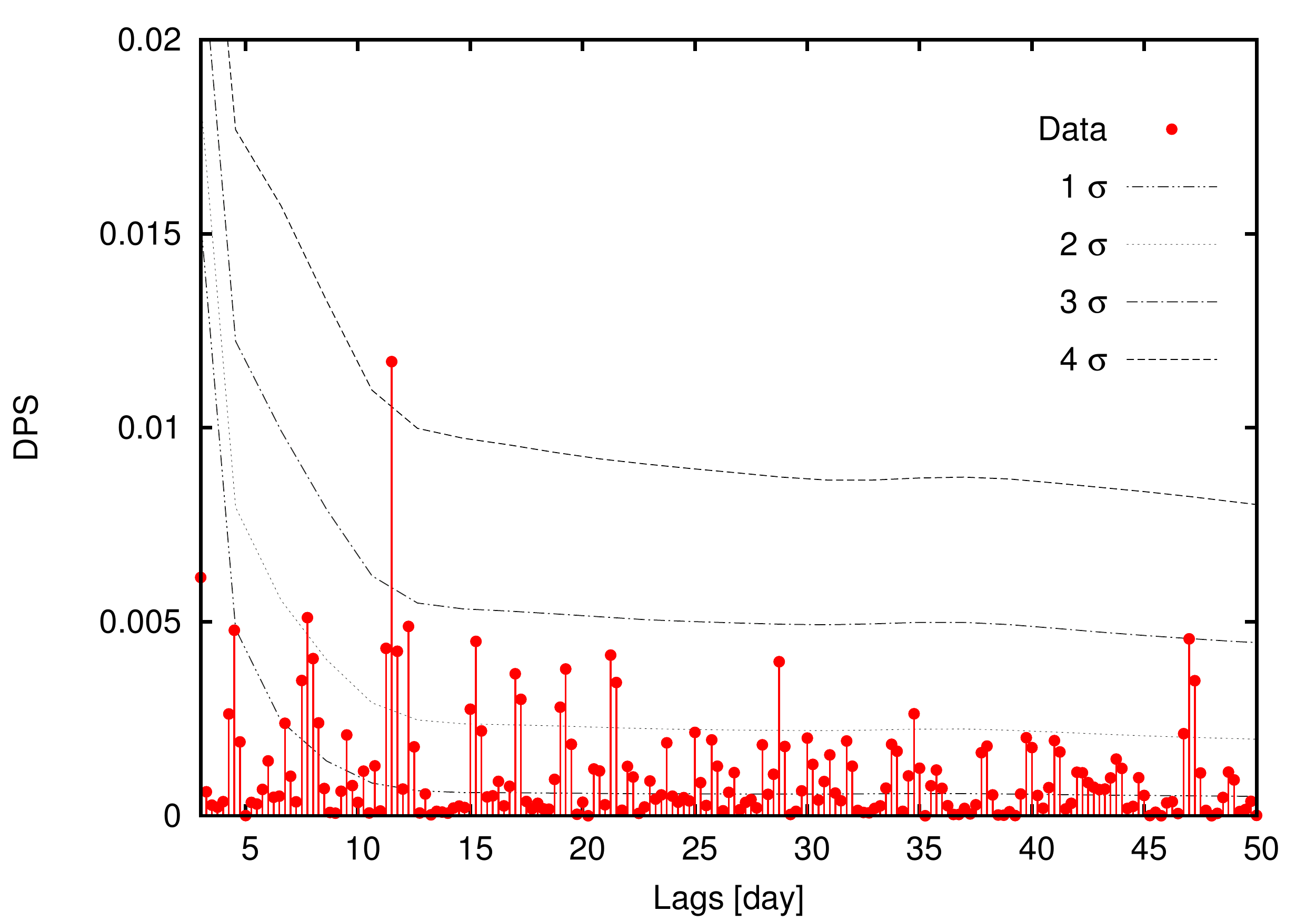}
\end{center}
\caption{\label{fig:dt_Flare1_all} 
			Autocorrelation Function ({\bf Left}) and Double Power Spectrum  ({\bf Right}) for Flare 1. % obtained from the sliced light curve. 
				 }
\end{figure*}

Enormous  gamma-ray activity occurred during the period MJD: 56160 -- 56280.
We analyze this time interval using the ACF and DPS (see Figure~\ref{fig:dt_Flare1_all}).
We obtain a time delay of $11.5\pm0.5\,$days for the ACF, and $11.38\pm0.13\,$days using the DPS. 

To estimate the significance of the detection, we use Monte Carlo simulations following \citet{2015ApJ...809..100B}.
We  produce $10^6$ artificial light curves.

\begin{figure*}
%\vskip 1cm
\begin{center}
\includegraphics[width=16cm,angle=0]{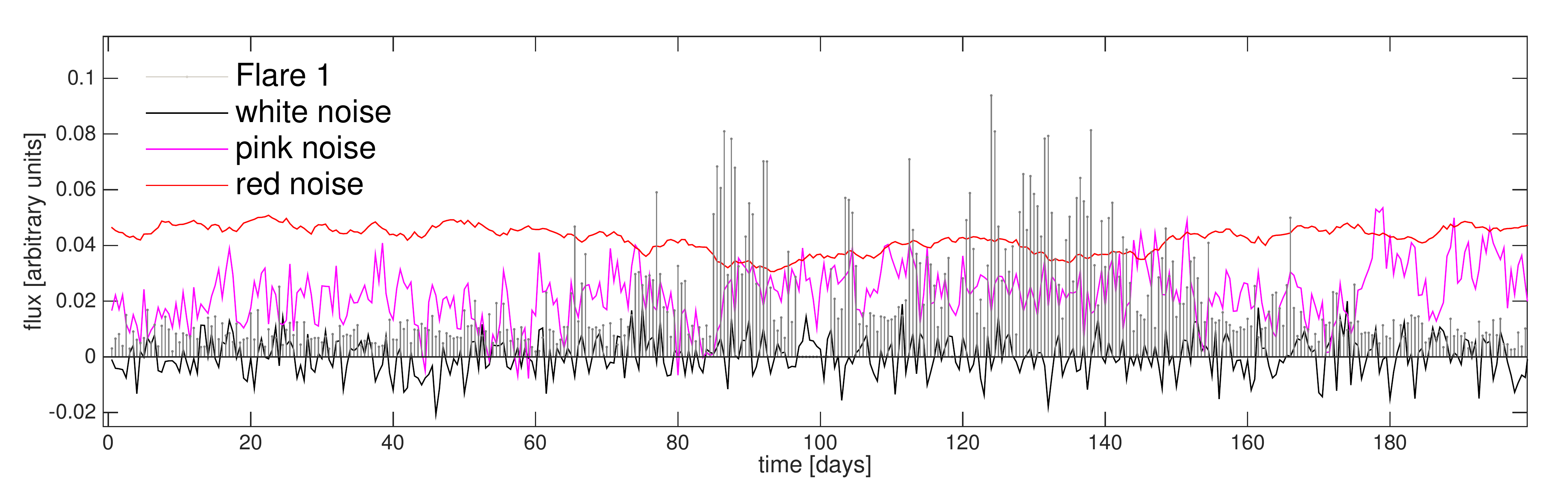}
\end{center}
\caption{\label{fig:MC_PLnoise} 
			Comparison of the  temporal behavior of Flare 1 and power law noise:
			white noise ($\alpha\sim0$),
			pink noise  ($\alpha\sim1$),
			and red noise  ($\alpha\sim2$). 
			We scale the flux of the Flare 1  by a factor of $2\times10^4$ to align the data with  the simulations.
			The flux of the pink and red noise are shifted by a factor of 0.02 and 0.04, respectively.
			The shift in scale  avoids overlapping of the plots,
			and facilitates  comparison of the temporal structure. 
				 }
\end{figure*}

The temporal behavior of Flare 1 is not  power law noise. 
The flare consists of a superposition of very short duration flares with large amplitudes.  
Figure~\ref{fig:MC_PLnoise}  compares  Flare 1 with the temporal behavior of power law noise with different indices.
The temporal behavior during Flare~1 is not well reproduced by any of these pure noise models. 
Pink noise ($\alpha \sim1$) is the closest match.
The power spectrum of  Flare 1 actually returns $\alpha = 0.9$.
However, pink noise is not  complete description of temporal behavior of Flare 1
because it cannot account for correlated bin-to-bin time variations of large amplitude. 
Increasing the index of power law noise (e.g. red noise)  
smooths the large fluctuations and obviously does not reproduce the behavior of the source. 
Decreasing the index toward  white noise increases the fluctuations on bin-to-bin time scales,
but the fast rise exponential decay profile that characterizes flares is absent \citep{2010ApJ...718..894P,2013ApJ...766L..11S}.
This correlated behavior resulting from the physics of the source is not reproduced by a simple noise model. 

Simulations of the signal composed of superpositions of short and bright flares are possible,
but the number of parameters required to define the flares is large. 
Thus following  \citet{2011A&A...528L...3B}, we divide  the light curve into  overlapping segments. 
We can use  this approach because  the flaring activity lasted for almost 200 days 
and the range of expected time delays is short, $<20\,$days.
We can thus divide the light curve of Flare~1 into  three overlapping segments of $128\,$days each.
We apply the ACF and DPS to each of the three segments and then average the results.
The error bars are the standard deviation among the three segments in each bin.

\begin{figure*}
%\vskip 1cm
\begin{center}
\includegraphics[width=8.cm,angle=0]{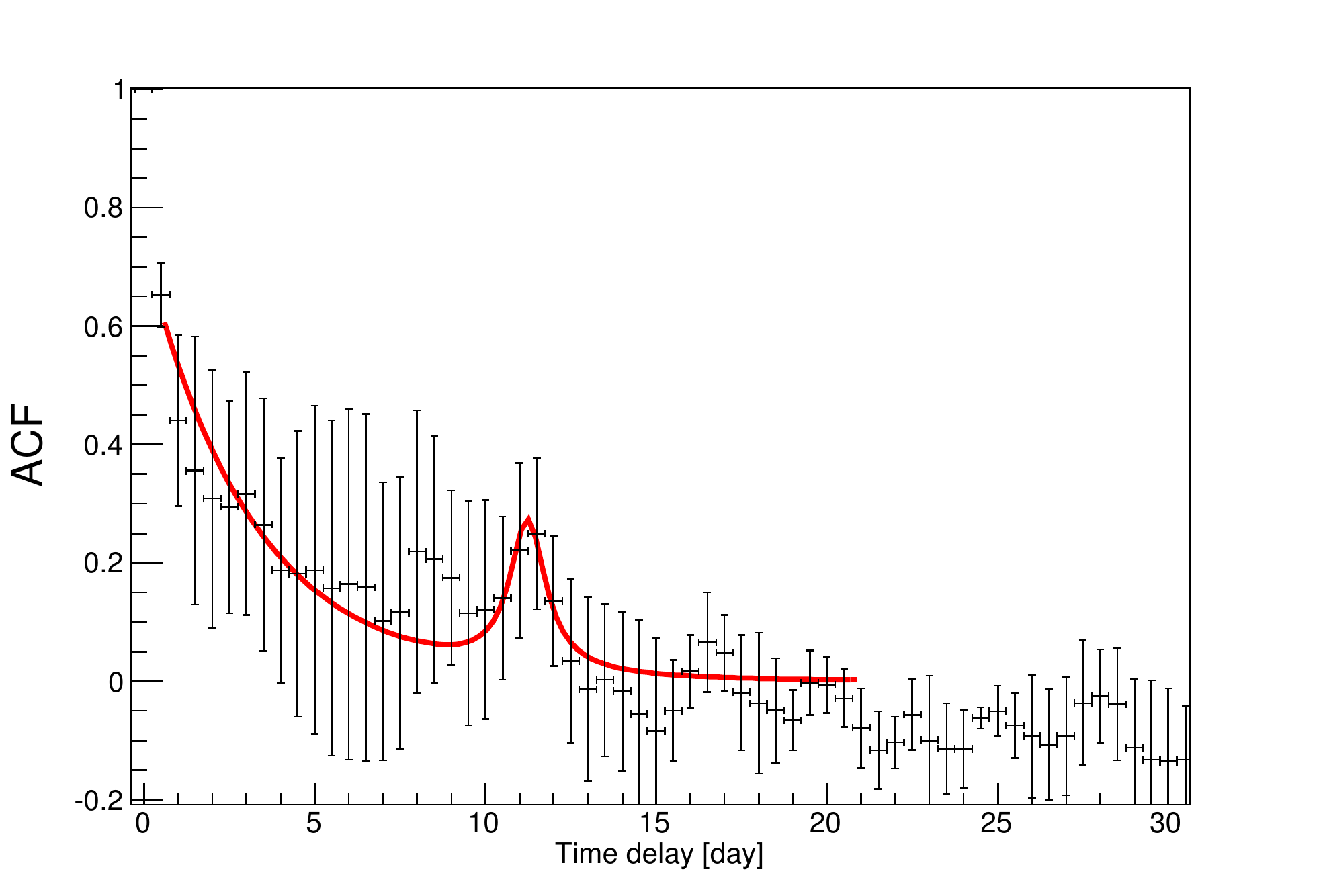}
\includegraphics[width=8.cm,angle=0]{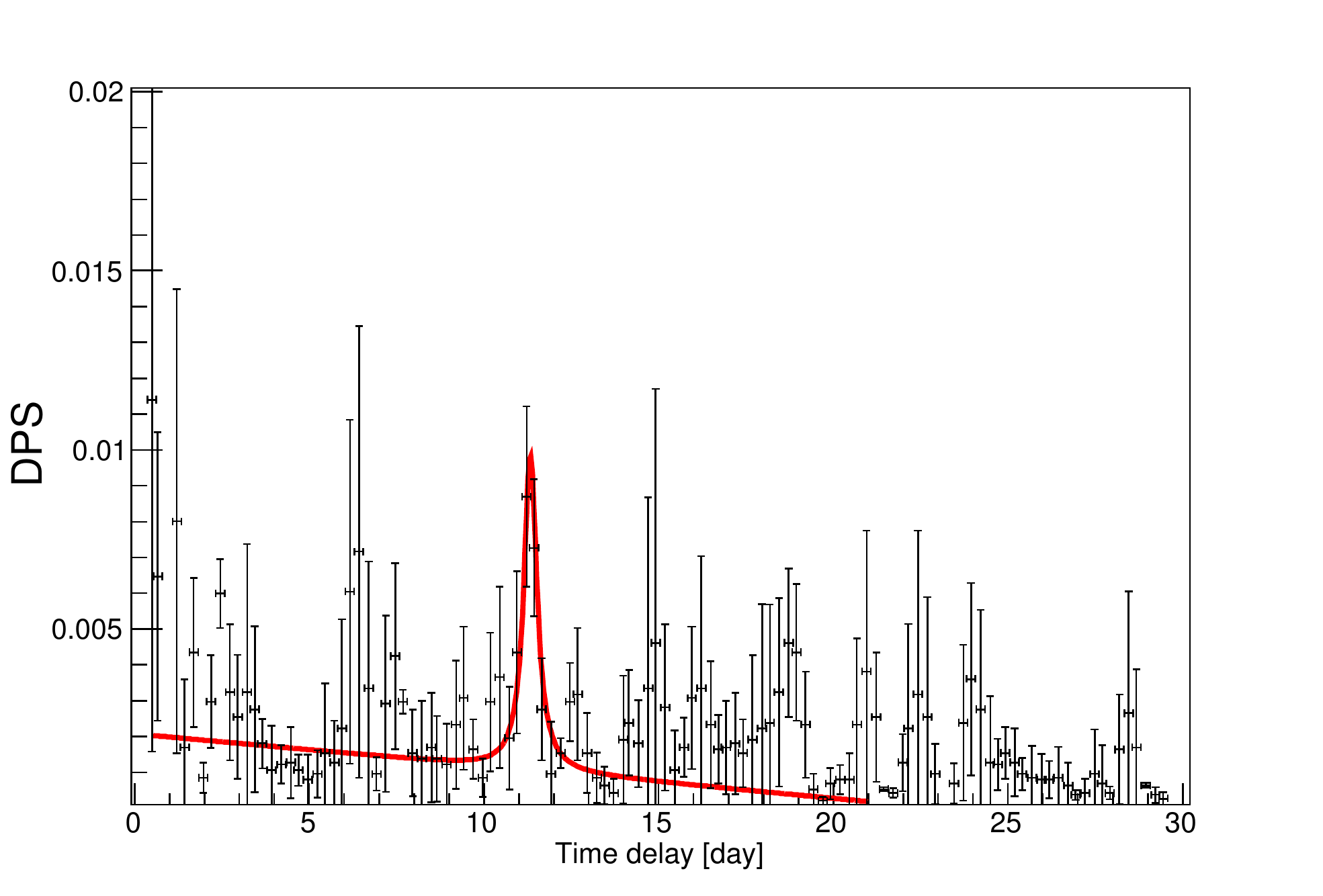}
\end{center}
\caption{\label{fig:dt_Flare1} 
			Autocorrelation Function ({\bf Left}) and Double Power Spectrum  ({\bf Right}) for Flare~1.
			The spectra are the average of three time sequences. 
			Solid lines are model fits, as described in the text.
				 }
\end{figure*}

Figure~\ref{fig:dt_Flare1} (Left) shows the ACF averaged over the segments.
We fit the spectrum with an exponential function representing the background
and  a Lorentzian function representing  the signal. 
The fit returns a time delay of $11.24\pm0.39\,$days.
However, the error bars are large and 
the significance of the signal is only $0.3\, \sigma$.

Figure~\ref{fig:dt_Flare1} (Right) shows the DPS from the averaged segments.
We fit linear plus Lorentzian profiles to obtain the position of the peak and its significance.
The corresponding time delay is $11.33\pm0.12\,$days. 
The signal is $4.13\,\sigma$ above the background. 
The DPS time delay, detected at high significance, agrees remarkably well  
with the previously obtained  gamma-ray time delay of $11.46\pm 0.16\,$days reported by \citet{HujCheung}.

The light curve is a superposition of multiple flares. 
The time between the random flares  could mimic a time delay.
These ``fake'' signals  should occur only in a fraction of the light curve.
A true gravitationally-induced time delay  persists over the entire flaring period. 
The analysis that averages over segments of the light curve  distinguishes the real, gravitationally induced, signal 
from randomly superimposed flares. 
If there is a real time delay the significance of the time delay increases with averaging over the three periods. 
For random multiple flares, the significance should not improve.   
In fact Figure~\ref{fig:dt_Flare1} shows that
multiple peaks are present because there is a lot of structure in the light curve.
However,  only the signal in bins around 11.5 days is significant (Figure~\ref{fig:dt_Flare1}).

%%%%%%%%%%%%%%%%%%%%%%%%%%%%%%%%%%%%%%
\subsection{Flare 2}
\label{sec:dt_Flare2}
%%%%%%%%%%%%%%%%%%%%%%%%%%%%%%%%%%%%%%

\begin{figure}
%\vskip 1cm
\begin{center}
\includegraphics[width=5.5cm,angle=-90]{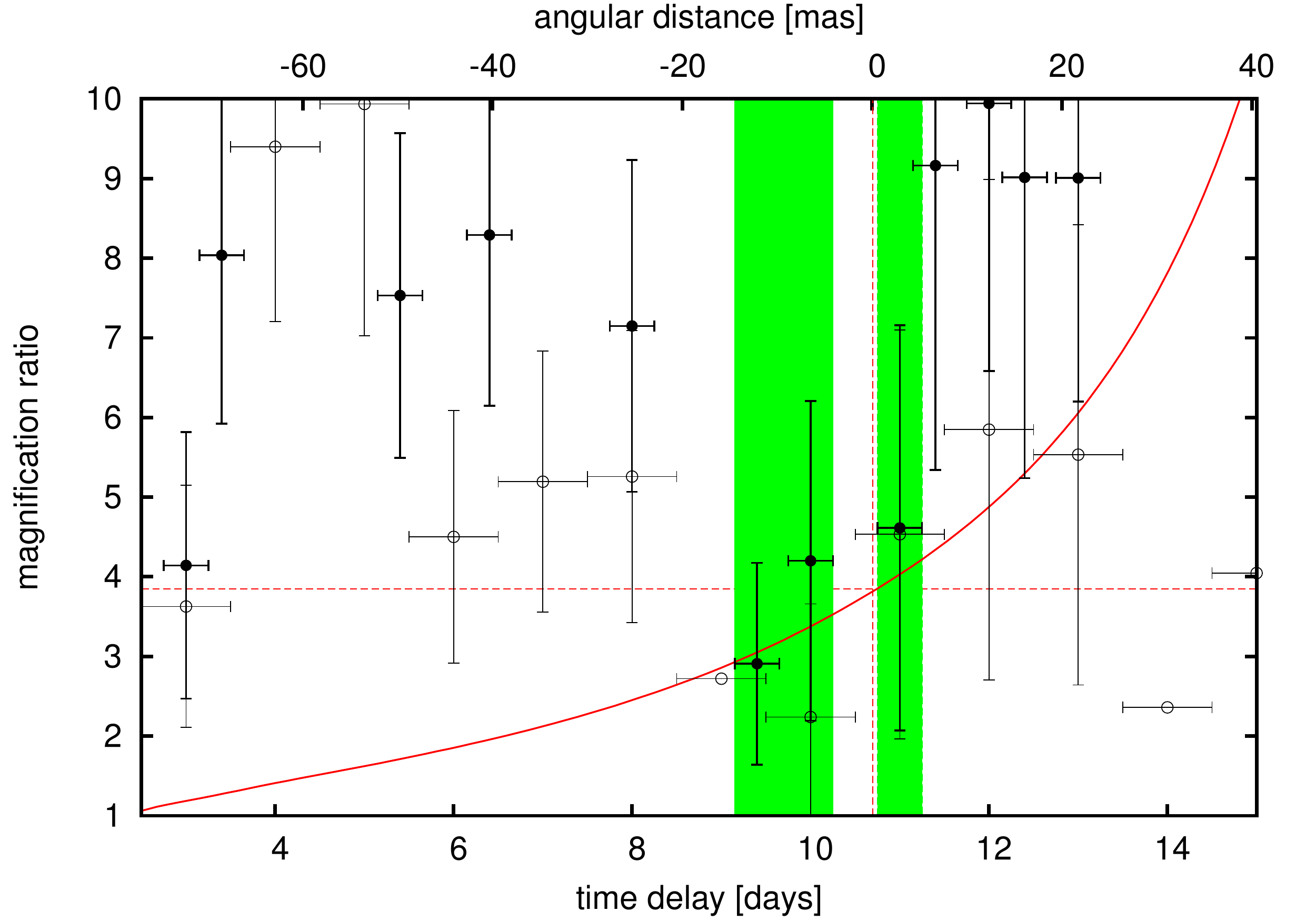}
\end{center}
\caption{\label{fig:2ndFlareMPM} 
                          The Maximum Peak Method applied to Flare~2.
                          Black and gray points are flux ratios calculated relative to the time bin center at MJD: 56852,
                          using light curve with 0.5~day and 1~day binning, respectively.
                          The solid red line indicates the predicted magnification ratio as a function of the time delay along the jet axis.
                          The red dotted lines indicates the value of the magnification ratio and time delay corresponding to the radio core resolved at 15~GHz.
                          The green area indicates values of time delay where the observed flux ratio is consisted with the model. 
                         }
\end{figure}

Flare 2, a single, bright flare, occurred in time period MJD: 56800 -- 57000.  
The light curve (2 day bins) around the flare consists mostly of upper limits (Figure~\ref{fig:lc_flares}).
This light curve is useless for  extracting several-day long time delays.
However, the huge advantage of having a single isolated flare is the ease of a direct search for the echo flare. 
Figure~\ref{fig:2ndFlareMPM} shows the result of application of
the Maximum Peak Method \citep[MPM, Section 3.2.3,][]{2015ApJ...809..100B}.
The MPM method suggests that the time delay lies in one of two ranges:  $9.75\pm0.5\,$days or $11\pm0.25\,$days. 
The errors corresponds to the bin width, not the $1\,\sigma$ standard deviation.

%%%%%%%%%%%%%%%%%%%%%%%%%%%%%%%%%%%%%%
\section{The Structure of the Gamma-ray Source}
\label{sec:HPT}
%%%%%%%%%%%%%%%%%%%%%%%%%%%%%%%%%%%%%%

So far we have used the radio observations and a lens model to reconstruct the origin of the radio core with a resolution of 1~mas  (Section~\ref{sec:LensResults}). 
The Fermi-LAT observations enable precise determination of  the time delay  for two gamma-ray flares (Section~\ref{sec:TimeDelay}). 
Here, we locate the sources of gamma-ray emission relative to the radio core by combining the radio source map 
and the Fermi-LAT time delays with the well-measured Hubble constant from \citet{2013arXiv1303.5076P}.

\citet{2015ApJ...799...48B} show that the Hubble parameter implied by the time delay is sensitive to any spatial  offset 
between the emission region that produces the resolved  mirage images and the site of the variable emission used to measure time delays. 
Purely on the basis of the physical processes involved, the gamma-ray emission from B2~0218+35 may not be spatially coincident with the radio core \citep{2014arXiv1403.5316B}.

The Hubble parameter, well measured with a variety of independent methods, provides a route to exploring this issue. 
We can use this precisely measured Hubble parameter to evaluate any offset 
between the radio core and the site of the variable gamma-ray emission.
We call this method the Hubble Parameter Tuning (HPT) approach. 

The Hubble parameter enters into the distance ratio in the time delay calculation (Eq~\ref{eq:dt}).
For an SIS gravitational potential, the relation reduces to:
\begin{equation}
\label{eq:h}
h = \frac{d(1+z_L)(\theta_B^2-\theta_A^2)}{2c\,\Delta t}\,.
\end{equation}

We have three kinds of constraints on the map of the source from radio to gamma-ray wavelengths:
the Hubble parameter, the positions of the lensed images, and the time delay between the images $\Delta t$.
If there is an offset between the radio core and the gamma-ray emitting regions, 
the Hubble parameter derived from the Fermi-LAT time delay will differ from the independently measured ``true'' value.
This difference depends on the distance between the radio core and the spatial location of the flare. 
The offset in Hubble space corresponds to the spatial offset  in the  source plane \citep{2015ApJ...799...48B}.

To locate the origin of the gamma-ray flares from B2~0218+35, 
we first  fix $\theta_A$ and $\theta_B$ to the  positions of the resolved images of the 15 GHz radio core (Table~\ref{tab:InputParameters}).
We use these image positions along with the model of the lens and the cosmological parameters
to infer the expected time delay for the position of the 15~GHz radio core.
Table~\ref{tab:Reconstruction} lists the reconstructed position;
the  value agrees well with the time delay  derived from the variability of the radio core although we do not use this delay to compute the Hubble constant.
The reconstructed time delay (Table~\ref{tab:Reconstruction})   
plugged  into  Equation~(\ref{eq:h}) is a consistency check which returns the true value of the Hubble parameter, our reference point.

Next, we calculate time delays for positions within $\sim10\,$mas from the radio core.
We use these time delays and positions of the 15 GHz images to compute the Hubble parameter using  Equation~(\ref{eq:h}). 
Figure~\ref{fig:HPT_Flare1} shows these calculated Hubble parameters as a function of the position of the variable emitting region. 
We call this projection of the model Hubble space.

%%%%%%%%%%%%%%%%%%%%%%%%%%%%%%%%%%%%%%
\subsection{The Spatial Origin of Flare 1}
%%%%%%%%%%%%%%%%%%%%%%%%%%%%%%%%%%%%%%

\begin{figure}
%\vskip 1cm
\begin{center}
\includegraphics[width=8.5cm,angle=0]{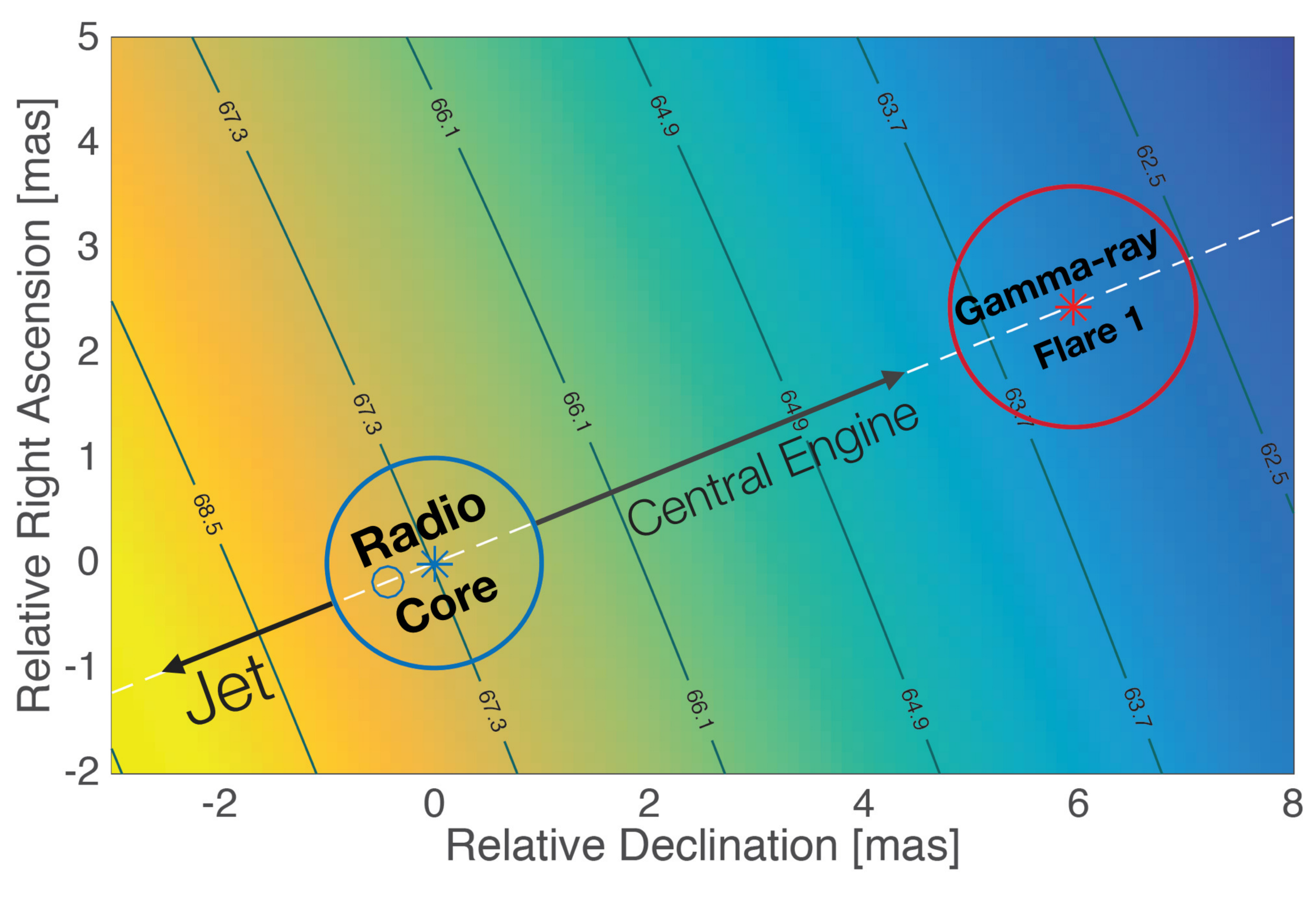}
\end{center}
\caption{\label{fig:HPT_Flare1} 
                          Hubble space.
                          The distances are shown with respect to the position of the radio core  (blue circle).
                          The radius of the blue circle corresponds to an uncertainty of $1\,$mas. 
                          The  blue star indicates the value of 
                          the Hubble parameter based on the  reconstructed position of the 15~GHz radio core. 
                          The open blue point shows the Hubble parameter derived from the observed  positions of the 15~GHz radio images.
                          The  dotted line shows the  jet projection. 
                          Gray arrows show the direction from the radio core toward the central engine and toward the jet. 
                          The red circle locates the spatial origin of Flare~1.
                          The radius of the red circle corresponds to the uncertainty in the time delay.  
                          The spacing of the white lines in Hubble space corresponds to $1.2$ km s$^{-1}$Mpc$^{-1}$, 
                          the  $1\,\sigma$ uncertainty in the Hubble parameter \citep{2013arXiv1303.5076P}.
                          }
\end{figure}

Flare~1 has a time delay of $11.33\pm0.12\,$days.
The Hubble parameter obtained based on the position of the 15 GHz radio core and this time delay corresponds to  $H_0 = 63.64 \pm 0.67\,\mbox{km\,s}^{-1}$Mpc$^{-1}$.
The quoted error  corresponds to an error in a time delay of $0.12\,$days
that translates into a spatial resolution of $1.15\,$mas.
Recall that H$_0=67.3\pm1.2\,\mbox{km\,s}^{-1}$Mpc$^{-1}$ from \citet{2013arXiv1303.5076P}.

We indicate the position of the 15 GHz radio core in Hubble space in Figure~\ref{fig:HPT_Flare1}.
The Hubble parameter estimated for Flare~1 appears as a red dot in Figure~\ref{fig:HPT_Flare1}.
The position of Flare~1 in Hubble space is displaced from the radio core. 
The resolved radio images also constrain the alignment  of the jet as indicated in  Figure~\ref{fig:HPT_Flare1} (white dotted line).

The distance between the 15~GHz core and the site of the gamma-ray flare is $6.4\pm1.1\,$mas displaced toward the central engine. 
This displacement corresponds to a projected distance of $51.2\pm8.8\,$pc.
The accuracy of the Hubble parameter measured with  \citet{2013arXiv1303.5076P},  $\pm1.2\,\mbox{km\,s}^{-1}$Mpc$^{-1}$, 
implies that the offset  between the resolved radio core and the variable gamma-ray site 
is significant at the $\sim3\,\sigma$ level. 
 
 %%%%%%%%%%%%%%%%%%%%%%%%%%%%%%%%%%%%%%
 \subsection{The Spatial Origin of Flare 2}
 %%%%%%%%%%%%%%%%%%%%%%%%%%%%%%%%%%%%%%

\begin{figure}
%\vskip 1cm
\begin{center}
\includegraphics[width=8.5cm,angle=0]{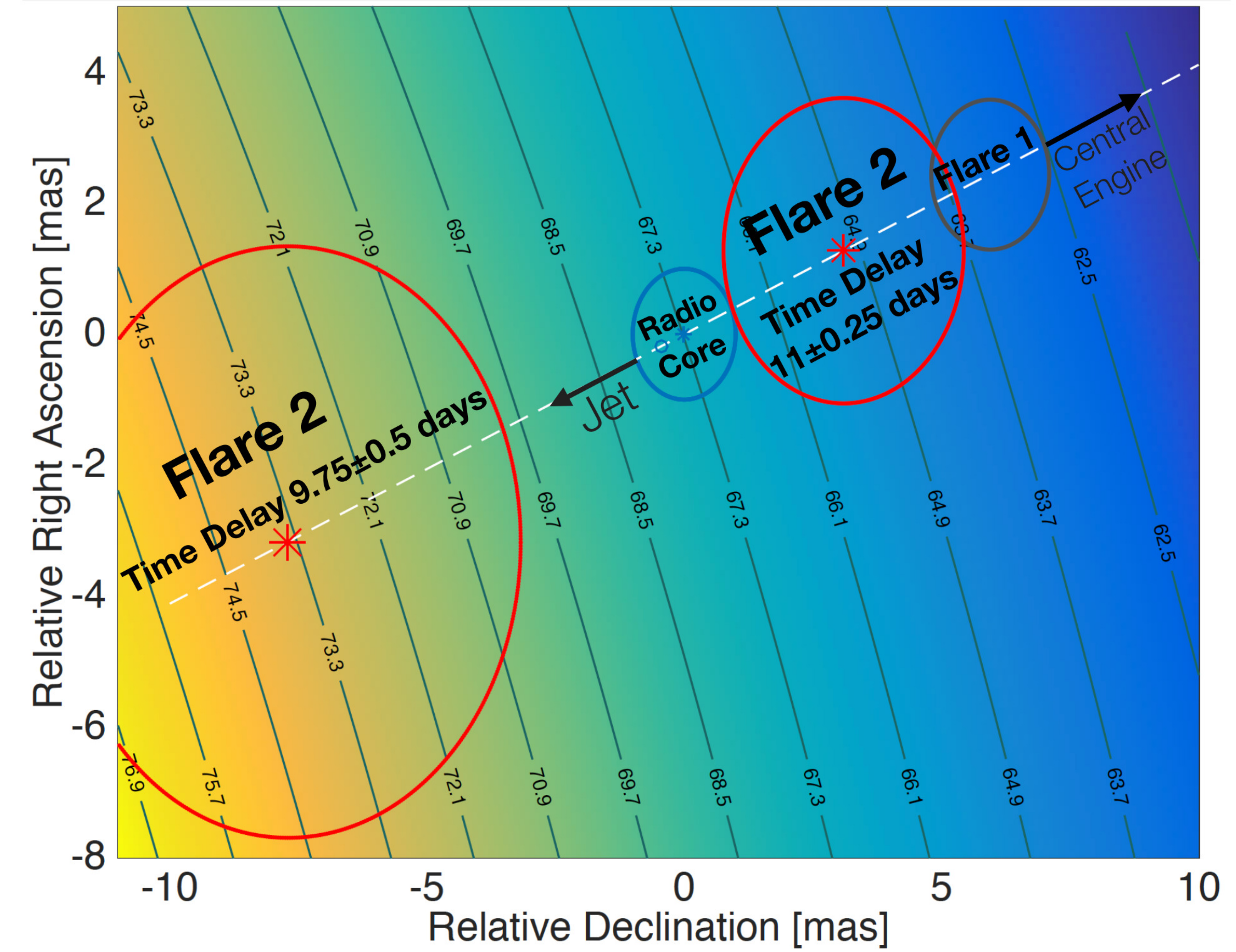}
\end{center}
\caption{\label{fig:HPT_Flare2} 
                          Hubble space calculated as in Figure~\ref{fig:HPT_Flare1}.
                          Red circles show the two possible sites for the  spatial origin of Flare 2 corresponding to
                          a time delay of $9.75\pm0.5\,$days or $11\pm0.25\,$days. }
\end{figure}

Flare~2 has a time delay  in one of two ranges:
$10.75-11.25\,$days or $9.25-10.25\,$days.
For the first range, the Hubble parameter is $63.64-66.6\,\mbox{km\,s}^{-1}$Mpc$^{-1}$.
The second range results in a Hubble parameter  of $69.85-77.4\,\mbox{km\,s}^{-1}$Mpc$^{-1}$.
We indicate the possible sites of Flare~2 in Figure~\ref{fig:HPT_Flare2}.
Flare~2 originates either $3.35\pm2.30\,$mas ($26.8\pm18.4\,$pc in the source plane) from the core toward the central engine,
or $8.33\pm4.5\,$mas ($66.64\pm36.00\,$pc in the source plane) in the direction of the jet. 

%%%%%%%%%%%%%%%%%%%%%%%%%%%%%%%%%%%%%%
\subsection{The Connection between Flare~1 and Flare~2}
%%%%%%%%%%%%%%%%%%%%%%%%%%%%%%%%%%%%%%

The position in the space of Hubble parameter versus offset shows  that Flare~2 is not coincident with either the core or Flare~1. 
Using the Hubble parameter tuning  approach, 
we can ask whether Flare~2 could result  from a  moving knot,
which first  produced Flare~1 and then moved downstream along  the  jet to produce Flare~2.  

The time between the beginning of Flare~1 and Flare~2, $\Delta t_{obs}$, is $690\,$days.
The projected distance between  Flare~1 and Flare~2, constrained by the time delay of $11\pm0.25\,$days, is  $D_{projected}\sim24\,$pc. 
In this case the model implies that knot is moving relativistically with an apparent velocity of $\beta_{app}$: 
\begin{equation}
\label{eq:projected}
\begin{split}
\beta_{app} & =  \frac{ D_{projected}(1+z_S)}{c\,\Delta t_{obs}}  \\
& \approx  70 \left(\frac{D_{projected}}{24\,\mbox{pc}}\right) \left(\frac{\Delta t_{obs}}{690\,\mbox{days}}\right)\,.
\end{split}
\end{equation}

Similar superluminal apparent motions of $\sim46\,c$ occur, for example, in the radio jet of the blazar PKS~1510-089 \citep{2005AJ....130.1418J}. 
Very high superluminal apparent motions  are commonly observed in gamma-ray blazar \citep{2013AJ....146..120L,2015ApJ...810L...9L}. 
This time delay thus yields a reasonable physical model for the gamma-ray source.

If the plasmon continues its motion with the same apparent velocity, $1.6\,$mas/year, 
it will pass through the stationary shock of  the 15~GHz core $\sim2\pm1\,$years after Flare~2, which was detected in July 2014.  
This model thus predicts increased radio emission in the time period around July 2016. 
Radio observations during this period could thus provide valuable insight into  the physical processes and plasma propagation along the jet. 

The second possible site of Flare~2, implied by the time delay of $\sim9.75\pm0.5$, is located at a projected distance of $\sim16\,$mas from Flare~1.
An apparent velocity of $350\,$c would be required to explain such a large projected distance.
Thus, these flares could not be produced by the same moving knot of plasma. 

We do not have direct evidence that Flare~2 is indeed connected with Flare~1.
However, the longer time delay implies a reasonable physical model for the source 
and  demonstrates the power of the Hubble parameter tuning approach.

%%%%%%%%%%%%%%%%%%%%%%%%%%%%%%%%%%%%%%
\section{Discussion}
\label{sec:Discussion}
%%%%%%%%%%%%%%%%%%%%%%%%%%%%%%%%%%%%%%

A major challenge of gamma-ray astronomy is localization of the emission region.
The blazar B2~0218+35 is uniquely suited to detailed reconstruction of the source position.
The gravitational lensing system is remarkably simple. 
There are exquisite radio data at several wavelengths along with the extensive Fermi-LAT light curve. 
These observations combined with the well-constrained Hubble constant enable 
the first reconstruction of the gamma-ray source positions relative to the radio core and jet.  

There are plausible sources of systematics that could in principle account for the 
the offset between the radio core and gamma-ray emission in B2~0218+35.
First, we consider a more complex mass distribution for the lens.
However, the time delay measured for the resolved radio core \citep[$10.5\pm0.4\,$days,][]{1999MNRAS.304..349B}
differs from the time delay measured for gamma-ray flares ($11.33\pm0.16\,$days). 
This difference clearly indicates that the complexity is in the source, not the lens. 
Even complex lens models cannot account for multiple time delays. 
Moreover, we have investigated a range of complex lens model.
More complex lens models were unable to reproduce the observations as well as the SIS model. 

Although the time delay difference rules out complexity of the lens, 
the accuracy of radio time delay (0.4 days) allows us to separate the radio core from the gamma-ray flare site only at the $\sim2\, \sigma$ level. 
The time delay can be translated into a relative position in the source plane. 
The accuracy of $0.4\,$days corresponds to $\sim3\,$mas. 
We use the position of the radio images reconstructed with $1\,$mas resolution to measure the distance between the emission sites ($51.2\pm8.8\,$pc), 
and we obtain a separation significant at the $\sim 3\, \sigma$ level.

We next consider systematics of the  Hubble constant measurement.
To measure the offset we use a Hubble parameter, H$_0=67.3\pm1.2\,$~km$\,$s$^{-1}\,$Mpc$^{-1}$, obtained by  \citet{2013arXiv1303.5076P}. 
Many independent methods provide a measure of H$_0$.
For example  the Hubble Space Telescope Key Project provides  H$_0=72\pm$8~$\mbox{km\,s}^{-1}$Mpc$^{-1}$ \citep{2001ApJ...553...47F}, 
the Cepheid distance ladder gives  $73.8\pm$2.4~$\mbox{km\,s}^{-1}$Mpc$^{-1}$ \citep{2011ApJ...730..119R,2011ApJ...732..129R}
and $74.3\pm1.5\,(\mbox{stat})\pm 2.1\,(\mbox{sys})\, \mbox{km\,s}^{-1}$ Mpc$^{-1}$ \citep{2012ApJ...758...24F}.

The gamma-ray time delay combined with the position of the radio core gives a Hubble parameter of  $H_0 = 63.64 \pm 0.67\,\mbox{km\,s}^{-1}$Mpc$^{-1}$.
Thus, even if we use the largest value ($74.3\pm1.5\,(\mbox{stat})\pm 2.1\,(\mbox{sys})\, \mbox{km\,s}^{-1}$ Mpc$^{-1}$) 
we obtain a significant offset between the radio core and the gamma ray of at least $\sim3\sigma$.  
Therefore, the separation between the radio core and the gamma-ray emission is robust to the large spread in values of the Hubble constant. 

Application of the method to other sources may not be as straightforward. 
For example, fewer constraints like well-resolved images together with time delays at the same frequency. 
Light curves may not be as well sampled as at gamma rays.
The radio data might be insufficient to reconstruct the projection of the jet.
Furthermore,  observations of relatively nearby sources show that jets can be bent, thus
introducing additional uncertainty  in measuring distances between emitting regions. 

%%%%%%%%%%%%%%%%%%%%%%%%%%%%%%%%%%%%%%
\subsection{Comparison with PKS1830-211}
%%%%%%%%%%%%%%%%%%%%%%%%%%%%%%%%%%%%%%

PKS~1830-211 is the only other gravitationally lensed  gamma-ray blazar known currently. 
Analysis of the gamma-ray time delays enable resolution of the origin of gamma-ray flares at the $\sim10\,$mas level, 
corresponding to $\sim100\,$pc in the source plane \citep{2015ApJ...809..100B}. 
In this case, the spatial resolution is limited by the accuracy of time delay, $\sim0.5\,$days.

Flare~1 of B2~0218+35 was  longer and brighter than the flares of  PKS~1830-211. 
The excellent photon statistics of Flare~1 allow measurement of the time delay with an accuracy of $\sim0.1\,$days, 
an improvement by a factor of 5 relative to PKS~1830-211. 

In PKS~1830-211 there is no Einstein ring. 
The pseudo-ring-like structure composed of images of the radio jet only 
allow  derivation of a boundary limiting the jet projection in the source plane. 

In contrast with PKS~1830-211, the well-resolved radio images and the more accurate gamma-ray time delay enable relative  localization of the radio and gamma-ray sources 
if the Hubble constant is sufficiently well-known from some independent technique. 
We can separate Flare~1 from the core in B2~0218+35 at the $\sim3\,\sigma$ level 
 because the Hubble parameter is measured with $\sim2\%$ accuracy 
 using the cosmic microwave background fluctuations \citep{2013arXiv1303.5076P}. 

 Future measurement of the Hubble parameter with an accuracy of $\sim1\%$ \citep{2014ApJ...794..135B}
 will allow  resolution of  the gamma-ray emission at even greater significance, $\sim6\,\sigma$.

%%%%%%%%%%%%%%%%%%%%%%%%%%%%%%%%%%%%%%
\subsection{The Spatial Origin of Gamma-ray Flares and Source Physics}
%%%%%%%%%%%%%%%%%%%%%%%%%%%%%%%%%%%%%%

The spatial origin of gamma-ray flares is a subject of debate \citep{2014ApJ...789..161N,2010MNRAS.405L..94T}.
Some observations indicate that flares are produced upstream from the resolved radio core \citep{2008Natur.452..966M,2010ApJ...710L.126M,2015arXiv151101085K}.
However, most  theoretical models assume that dissipation takes place within a sub-parsec  scale from the central engine, 
where the plasma is denser and the sources of seed photons for inverse-Compton radiation are abundant  
\citep{2014ApJ...796L...5N,2014A&A...567A.113B,2015MNRAS.448.3121H,2011ApJ...733...19T}.  

Only radio telescopes can resolve scales smaller than 100~pc.
Radio telescopes  probe synchrotron radiation produced in optically thick parts of the jet. 
The radio core is often interpreted as  the location where the jet opacity to synchrotron self-absorption is 1 \citep{1979ApJ...232...34B}.
Observationally,  the radio core is the region of peak intensity generated by a compact component at the apparently upstream end of the jet  \citep{2008ASPC..386..437M,2015ApJ...807...15H}. 

Measurement of the distance between a radio core and the central engine is difficult. 
These  measurements have only been possible for nearby sources with prominent two-sided jets.
Observing outflows in two opposite directions then constrains the position of the central engine  \citep{2015ApJ...807...15H}.

For gamma-ray blazars the two-sided jet is presumably present, 
but the counterjet is too faint for detection because of relativistic beaming.
Thus, in blazars, we observe only one sided-jets. 
As summarized by \citet{2015ApJ...807...15H}, the general picture is  that a central engine exists somewhere upstream of observed cores. 
B2~0218+35 provides the first direct observational constraint.

In B2~0218+35 the bright gamma-ray flare occurred upstream in the jet at the projected distance of $51.2\pm8.8\,$pc from the 15~GHz radio core.
Thus the central engine must be at least this far away from the radio core. 
The Hubble tuning approach we use can enable measurement of the apparent velocities of gamma-ray emitting knots. 
Based on this measurement, we can predict the time of a plausible interaction with the radio core. 
For B2~0218+35 we predict increased emission at gamma ray and/or radio wavelengths   within a year of July 2016. 

%%%%%%%%%%%%%%%%%%%%%%%%%%%%%%%%%%%%%%
\subsection{Gravitational Lensing and  X-ray Sources}
%%%%%%%%%%%%%%%%%%%%%%%%%%%%%%%%%%%%%%

The improvement of angular resolution at X-ray wavelengths with {\it Chandra} enabled the discovery
of  high energy extragalactic jets extending over hundreds of kiloparsecs 
\citep{2000ApJ...540L..69S,2000ApJ...542..655C,2002ApJ...570..543S,2002ApJ...571..206S,2004ApJ...608..698S,2005ApJS..156...13M,2006ApJ...641..717S,
2006ARA&A..44..463H,2007ApJ...662..900T}
%1991AJ....101.1632B}.
In fact, our investigations of the structure of lensed gamma-ray sources were inspired by {\it Chandra}'s discovery of flaring emission  from HST-1, 
a knot of strong X-ray emission displaced from the core in the nearby galaxy M87 \citep{2006ApJ...640..211H}.
Recently, deep {\it Chandra} observations of Pictor A revealed the high-energy  flares located in knots displaced along the jet \citep{2015arXiv151008392H}.

For  sources at redshift $\sim1$, the {\it Chandra} resolution of 0.5~arcseconds corresponds to 4~kpc.    
Thus, knots like HST-1 cannot be resolved.
Improvement in angular resolution of at least two orders of magnitude is needed  to resolve these sources and to  explore their  evolution. 
 
There are, however, $\sim20$ gravitationally lensed quasars with associated X-ray emission. 
The sources have been monitored in  searches for  microlensing \citep{2002ApJ...568..509C,2003ApJ...589..100D,2012ApJ...757..137C,2012ApJ...755...24C,2013ApJ...769...53M}.
These data and future monitoring of variable X-ray sources may enable   measurement of time delays. 

These sources have been also monitored at radio and optical wavelengths. 
Combining these observations along the lines we have followed for B2~0218+35  may enable reconstruction of these sources based on lensing models. 
If displacements of the X-ray emitting regions along the jets are common, they pose challenges to understanding of the particle acceleration mechanism. 
They may also increase our understanding of the distribution of Hubble constants derived from time delays \citep{2015ApJ...799...48B}.

%%%%%%%%%%%%%%%%%%%%%%%%%%%%%%%%%%%%%%
\subsection{Gravitational Lensing and SKA}
%%%%%%%%%%%%%%%%%%%%%%%%%%%%%%%%%%%%%%

SKA will observe thousands of gravitationally lensed quasars 
with a resolution of $\sim2\,$mas at $10\,$GHz,
and $\sim20\,$mas at $1\,$GHz  \citep{2009IEEEP..97.1482D,2012PASA...29...42G,2015aska.confE..84M}.
These radio observations will provide a foundation for reconstructing the mass distribution of lenses and the positions of radio cores. 

All  radio quasars have X-ray emission \citep{2015eheu.conf...26R}.
Thus, among these lensed systems, there will  be a  large population of quasars  with  variable high energy emission where time delays can be measured. 

An ensemble of high-energy quasars with measured time delays and reconstructed source structure based on robust lensing models will enable
investigation  of the origin of X-ray radiation  and the connection between the radio and high-energy emission.
The  large redshift range of lensed quasars  should provide constraints on the co-evolution of radio and  high-energy jets. 

%%%%%%%%%%%%%%%%%%%%%%%%%%%%%%%%%%%%%%
\section{Conclusions}
\label{sec:Conclusions}
%%%%%%%%%%%%%%%%%%%%%%%%%%%%%%%%%%%%%%

B2~0218+35 is one of only two known gravitationally lensed systems detected at  gamma rays.
We reconstruct the mass distribution of its lensing galaxy and the properties of the jet based on  well-resolved radio observations. 
We use the dense Fermi-LAT light curves to measure time delays for  two gamma-ray flares. 
The  position of the mirage images, the time delay, and the independently known Hubble constant enable measurement of
the spatial offset between the gamma-ray emission region and the radio core.

Our reconstruction of the lensed source shows that the extended flare (Flare 1) is displaced 
from the radio core at the $\sim 3\sigma$ level. 
The displacement is upstream from the jet providing the first direct observational constraint on the location of the  central engine relative  to the radio core  for blazars. 

A shorter flare (Flare~2) may be an event following Flare~1. 
If  Flare~1 and Flare~2 are indeed  connected  then  the knot which produced these gamma-ray flares moves at apparent velocity of $\sim70\,$c. 
Thus the model  makes a testable prediction. 
There should be an interaction with the radio core  within a year of July 2016. 

The only other known gravitationally lensed gamma-ray blazar is PKS~1830-211.
This source also has complex structure 
with gamma-ray flares originating from multiple regions along the jet
\citep{2011A&A...528L...3B,2015ApJ...809..100B}.   

Lensed high-energy sources  monitored with detectors like {\it Chandra}, Swift and NuSTAR
offer rich opportunities to extend this powerful lens modeling approach  to other sources. 
There are more that 20  known lensed quasars with associated X-ray emission. 
Some of these systems already have enough observations to reconstruct the mass distribution of the lens. 
Further monitoring will enable measurement of time delays with X-ray detectors. 

In the near future, SKA will resolve thousands of radio images of gravitationally lensed quasars.
Many of these quasars will also have prominent variable  X-ray emission, allowing recovery of time delays. 
This ensemble of observations combined with the power of strong gravitational lensing 
will probe the origin of X-ray radiation, its connection to radio emission, and  the cosmic evolution of jets at radio and high energies.

%%%%%%%%%%%%%%%%%%%%%%%%%%%%%%%%%%%%%%
\acknowledgments
%%%%%%%%%%%%%%%%%%%%%%%%%%%%%%%%%%%%%%
We thank the referee for providing valuable comments that helped us to improve the manuscript and have led to adding the appendix.
We thank Scott Kenyon and Dan Schwartz for  valuable comments. 

A.B. is  supported by NASA through Einstein Postdoctoral Fellowship.
MJG is supported by the Smithsonian Institution.
%We thank the referee for valuable comments on the manuscript.

%%%%%%%%%%%%%%%%%%%%%%%%%%%%%%%%%%%%%%
\appendix
\section{Monte Carlo Simulations and Lens Modeling}
\label{app}
%%%%%%%%%%%%%%%%%%%%%%%%%%%%%%%%%%%%%%

Selecting the class of models is a key part of lens modeling.
A simple and useful mass model of a galaxy acting as a lens is an isothermal model with density $\rho \propto r^{-2}$ and a flat rotation curve. 
Moreover, the isothermal profile is consistent with a large range of observations of spiral and elliptical galaxies 
\citep{1989ARA&A..27...87F,1993ApJ...416..425M,1995ApJ...445..559K,1997ApJ...488..702R,2001ApJ...554.1216C,2002ApJ...575...87T,2003ApJ...583..606K}. 

Here, our goal is to find a unique mass distribution for the lens. 
Therefore, we investigate  a lens profile allowing for a more complex mass distribution.
We use a softened power law potential \citep[][Eq. 25]{2001astro.ph..2341K}: 
\begin{equation}
\label{eq:modelMC}
\phi = b(s^2+x^2+y^2/q^2)^{\alpha/2} - b\, s^\alpha \,,
\end{equation}
where
$b$ is the Einstein radius,
$s$ is a scale radius of a flat core (in mas),
$q$ is the projected axis ratio, and $\alpha$ is a power law exponent. 
The softened power law potential with  $s=0$, $q=1$, and $\alpha=1$ 
reduces to  the SIS potential.

Table~\ref{tab:Range} shows the range of parameters we  explore using Monte Carlo simulations. 
These range of parameters are constrained based on the set of radio and optical observations described in Section~\ref{sec:conlens}.

%---------------------- Table: Range-------------------------------%

\begin{table}[h!]
  \begin{center}
    \caption{Range of Parameters for the Monte Carlo Simulations}
    \label{tab:Range}
    \begin{tabular}{ccc}
   \hline
      Parameter &  Min & Max \\
   \hline
      $s$ [mas] & 0 & 50 \\
      $q$ & 0.9 & 1 \\
      $\alpha$  & 0.96 & 1.04 \\
      $x_S$ & 350 &  370 \\
      $y_S$ & 332.5 & 352.5 \\
      $x_L$ & 248 &  288 \\
      $y_L$ & 284.6 & 324.6 \\
   \hline
    \end{tabular}
  \end{center}
\end{table}

%--------------------------------------------------------------------%

\begin{figure*}
%\vskip 1cm
\begin{center}
\includegraphics[width=17.cm,angle=0]{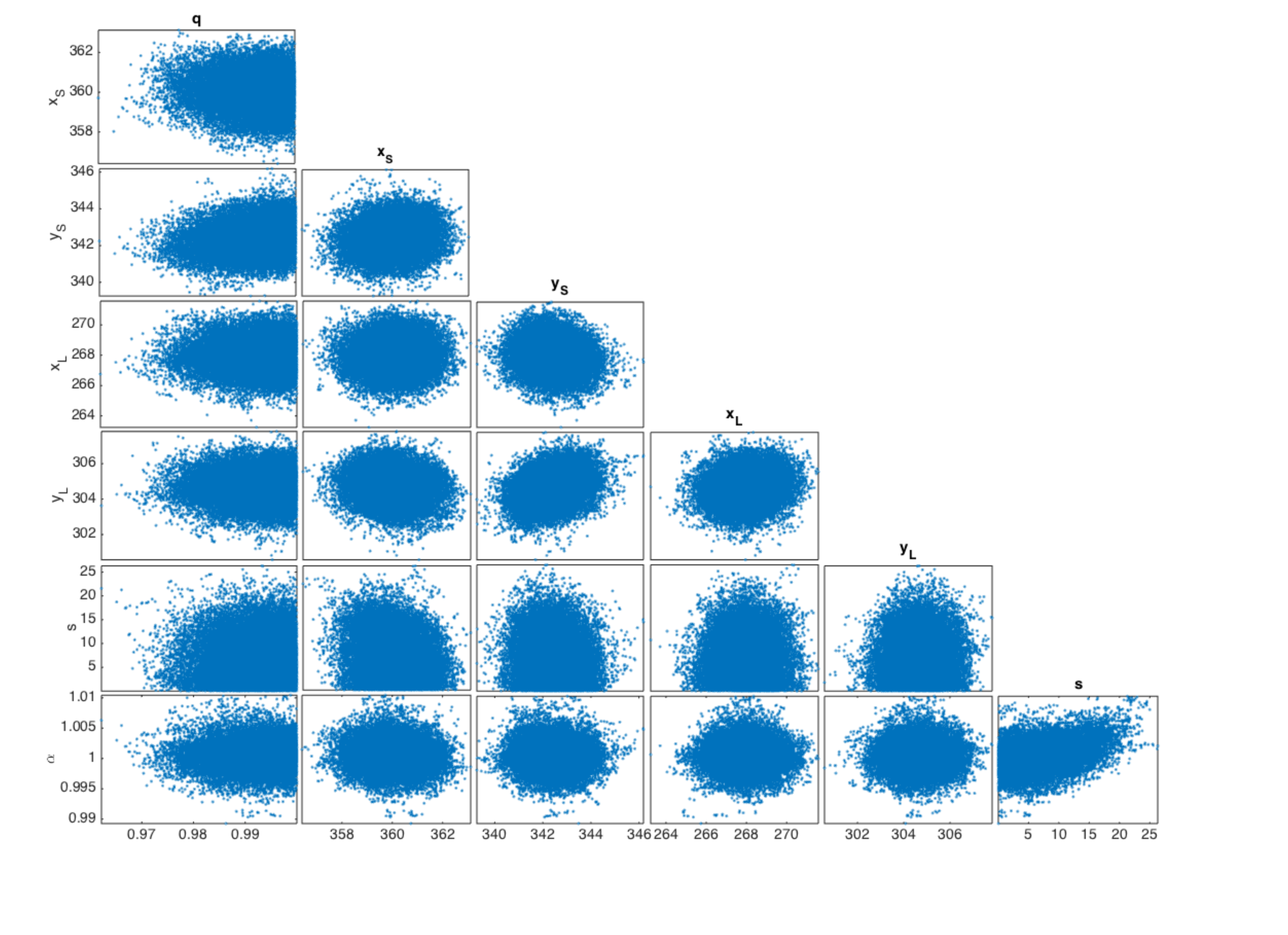}
\end{center}
\caption{\label{fig:MC1} 
                            Pairwise scatterplots showing the bivariate marginal distributions for  a softened power law potential 
			simulated using the Metropolis-Hastings chains.
				 }
\end{figure*}

\begin{figure*}
%\vskip 1cm
\begin{center}
\includegraphics[width=16cm,angle=0]{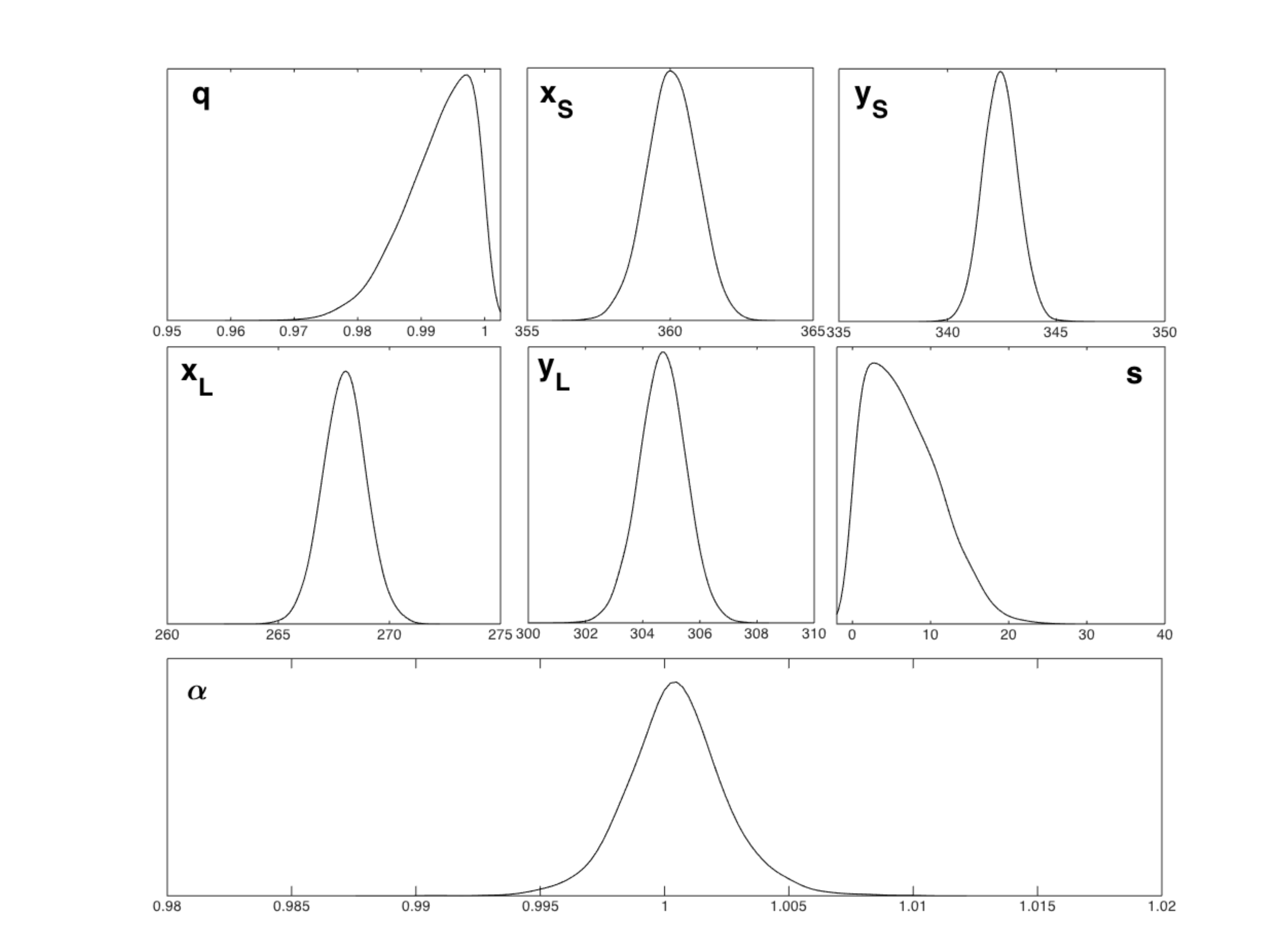}
\end{center}
\caption{\label{fig:MC2} 
			The univariate marginalized probabilities for all parameters of the softened power law potential.
				 }
\end{figure*}

We use Monte Carlo simulations to explore lens models defined by Eq. (\ref{eq:modelMC}) 
varying in the range of parameters listed in Table~\ref{tab:Range}.
We use the MCMC\footnote{http://helios.fmi.fi/$\sim$lainema/mcmc/} toolbox within {\tt MATLAB}.
This toolbox generates and analyzes  Metropolis-Hastings MCMC chains using a multivariate Gaussian proposal distribution.
Table~\ref{tab:MCResults} and Figures~\ref{fig:MC1} and \ref{fig:MC2}   shows the results of the MC simulations. 
The derived  parameters are very close to the ideal SIS where $s=0$, $q=1$, and $\alpha=1$.

%---------------------- Table: MC Results-------------------------------%

\begin{table}[h!]
  \begin{center}
    \caption{Results of the Monte Carlo simulations.  }
    \label{tab:MCResults}
    \begin{tabular}{cccccc}
   \hline
      Parameter &  mean\footnote{The mean values estimated from the chain of $5\times 10^5$ simulations} & std\footnote{The standard deviations} & 
      $MCerr$\footnote{An estimate of the Monte Carlo error} & $\tau$\footnote{The integrated autocorrelation time} & $geweke$\footnote{A simple test for a null hypothesis that the chain has converged} \\
   \hline
         $s$  & 6.5639 &   4.5181 &  0.1014     & 28.90 &   0.971 \\
         $q$  & 0.9925 &  0.0055  &  0.0001     & 25.19 &  0.999 \\
$\alpha$  & 1.0005 &  0.0020  &  5.08e-05 & 29.60 &  0.999 \\ 
  $x_S$    & 360.06 &  0.8658  &  0.0205     & 29.63 &  0.999 \\
  $y_S$    & 342.44 &  0.7878  &  0.0174     & 29.18 &  0.999 \\
  $x_L$    & 267.98 &  0.9659  &  0.0203     & 23.79 &  0.999 \\
  $y_L$    & 304.69 &  0.8072  &  0.0178     & 27.11 &  0.999 \\
   \hline
    \end{tabular}
  \end{center}
\end{table}

%--------------------------------------------------------------------%

The Monte Carlo simulations show that the best global solution for the lens model for B2~0218+35 is the SIS. 
To exclude other solutions, we have run an additional  $10^6$ MC simulation uniformly sampled over the range of parameters from Table~\ref{tab:Range}.
We identified  local minima with parameters for the lens potential that reproduce the position of the images. 
All models that reconstruct the image  within at least 5 mas are very close the SIS. 
Moreover, all models with a more complex gravitational potential  that provide a reasonable reconstruction of the mirage images fail to reproduce the radio time delay. 
%We where able to reject does models at more than 3$\,\sigma$ level using the radio time delay.

Precise reconstruction of the distance between the emitting regions of  the well resolved mirage images,
and the emitting region with the well measured time delay rely  on knowledge of the mass distribution of the lens. 
Different lens models result in different rates of change of the time delay as a function of the position of the source. 
In principle, this rate is the source of systematics in calibrating the distance between the emitting regions. 

We calculate the rates of change in the time delay around the source position for different parameters of the lens model. 
We calculate the change in the time delay for source positions moved by $6\,$mas in the radial direction. 
We calculate the expected change in the time delay for models deviating from the SIS. 
We use the error in the lens parameters estimated from the Monte Carlo simulations (Table~\ref{tab:MCResults}). 
We investigate lens models with parameters $\alpha$, $s$, and $q$ within $1\,\sigma$ and $3\,\sigma$  from the SIS.  

\begin{table}[h!]
  \begin{center}
    \caption{}
    \label{tab:Rate}
    \begin{tabular}{cccccccc}
   \hline
      Parameter within $1\, \sigma$ &  $d\Delta t / dr$ & $d Ratio/ dr$  & Dist [mas] &  Parameter within $3\, \sigma$ &  $d\Delta t / dr$ & $d Ratio/ dr$  & Dist [mas]  \\
   \hline
             SIS & $0.61 \, \mbox{days} / 6 \, \mbox{mas}$ & $ 0.55 / 6 \, \mbox{mas}$ & 1.25 &             SIS & $0.61 \, \mbox{days} / 6 \, \mbox{mas}$ & $ 0.55 / 6 \, \mbox{mas}$  \\
     \hline
     $\alpha = 0.998$ & $0.613 \, \mbox{days} / 6 \, \mbox{mas}$ & $ 0.47 / 6 \, \mbox{mas}$ & 4.5 &  $\alpha = 0.994$ & $0.62 \, \mbox{days} / 6 \, \mbox{mas}$ & $ 0.48 / 6 \, \mbox{mas}$ & 12  \\
     $\alpha = 1.002$ & $0.61 \, \mbox{days} / 6 \, \mbox{mas}$ & $ 0.5 / 6 \, \mbox{mas}$ & 4.6  &    $\alpha = 1.006$ & $0.6 \, \mbox{days} / 6 \, \mbox{mas}$ & $ 0.48 / 6 \, \mbox{mas}$ & 12   \\
         $s = 4.5$ & $0.61 \, \mbox{days} / 6 \, \mbox{mas}$ & $ 0.55 / 6 \, \mbox{mas}$ & 1.8            &    $s = 13.5$ & $0.61 \, \mbox{days} / 6 \, \mbox{mas}$ & $ 0.55 / 6 \, \mbox{mas}$ & 1.8    \\
          $q = 1.0055$ & $0.58 \, \mbox{days} / 6 \, \mbox{mas}$ & $ 0.5 / 6 \, \mbox{mas}$ & 2.6             & $q = 1.0165$ & $0.7 \, \mbox{days} / 6 \, \mbox{mas}$ & $ 0.43 / 6 \, \mbox{mas}$ & 4.6   \\
   \hline
    \end{tabular}
  \end{center}
\end{table}

Table~\ref{tab:Rate} shows the expected differences in time delays and magnification ratios. 
We also list the resulting difference between the reconstructed position of the images and positions of the observed images.  
For example, even a change as small as 0.002 in the $\alpha$ parameter results in a significant change in the reconstructed image positions.
The calculations summarized in  Table~\ref{tab:Rate}  demonstrate that the estimated offset between the radio core and region of the gamma-ray flare of $6.6\pm1.1\,$mas does not suffer from systematics in the lens modeling. 
The offset is estimated with  $\sim 20\%$  accuracy. 
As demonstrated in Table~\ref{tab:Rate},
the lens model cannot introduce  uncertainties of more that 5\%. 
Therefore, our estimation of the offset between the radio core and gamma-ray flare is robust.

\bibliography{B20218_gamma}
\end{document}